\documentclass[usenatbib]{mnras}
\usepackage{aas_macros}
\usepackage{amsmath}
\usepackage{amssymb}
\usepackage{color}
\usepackage{epsfig}
\usepackage{float}
\usepackage{graphicx}
\usepackage{latexsym}
\usepackage{morefloats}
\usepackage{natbib}
\usepackage{subfigure}
\usepackage{times}
\usepackage{multirow}

\newcommand{\plotone}[1]{\resizebox{0.95\hsize}{!}{\includegraphics{#1}}}

\newcommand{\plotthree}[3]{\center {\resizebox{0.95\hsize}{!}
{\includegraphics{#1}\hspace{0.5cm}\includegraphics{#2}\hspace{0.5cm}\includegraphics{#3}}}}

\newcommand{\Msun}{{~\rm M_\odot}}

\newcommand{\kpc}{~\rm kpc}
\newcommand{\Mpc}{~\rm Mpc}

\newcommand{\uvec}[1]{\boldsymbol{\mathit{\hat{#1}}}}

\def\gsim { \lower .75ex \hbox{$\sim$} \llap{\raise .27ex \hbox{$>$}}}
\def\lsim { \lower .75ex \hbox{$\sim$} \llap{\raise .27ex \hbox{$<$}}}
\newcommand{\eagle}{\textsc{eagle}}
\newcommand{\apo}{\textsc{apostle}}
\newcommand{\auriga}{\textsc{auriga}}
\newcommand{\healpix}{\textsc{HEALPix}}
\newcommand{\simRef}{Ref-L{\small 0100}N{\small 1504}}
\newcommand{\lcdm}{$\Lambda$CDM}

\newcommand{\refsec}[1]{Sec. \ref{#1}}

\newcommand{\reffig}[1]{Fig. \ref{#1}}

\title[Galactic satellite accretion]
{The multiplicity and anisotropy of galactic satellite accretion}

\author[Shi Shao et al.]
{\parbox{\textwidth}{
Shi Shao$^{1}$\thanks{E-mail: shi.shao@durham.ac.uk}, Marius
Cautun$^{1}$, Carlos~S.~Frenk$^{1}$, Robert J. J. Grand$^{2,3}$, \\
Facundo A. G\'{o}mez$^{4,5}$ Federico Marinacci$^{6}$ and
Christine~M.~Simpson$^{2}$
 \vspace{.20cm}} \\
$^1$Institute for Computational Cosmology, Department of Physics,
Durham University, South Road Durham DH1 3LE, UK \\
$^2$Heidelberger Institut f\"{u}r Theoretische Studien,
Schloss-Wolfsbrunnenweg 35, 69118 Heidelberg, Germany \\
$^3$Zentrum f\"{u}r Astronomie der Universit\"{a}t Heidelberg, ARI,
M\"{o}nchhofstr. 12-14, 69120 Heidelberg, Germany \\
$^{4}$Instituto de Investigaci{\'o}n Multidisciplinar en Ciencia y
Tecnolog{\'i}a, Universidad de La Serena, Ra{\'u}l Bitr{\'a}n 1305, La Serena, Chile \\
$^{5}$Departamento de F{\'i}sica y Astronom{\'i}a, Universidad de La
Serena, Av. Juan Cisternas 1200 N, La Serena, Chile \\
$^6$Department of Physics, Kavli Institute for Astrophysics and Space Research,
MIT, Cambridge, MA 02139, USA \\
}

\begin{document}
\maketitle
\begin{abstract}
We study the incidence of group and filamentary dwarf galaxy accretion
into Milky Way (MW) mass haloes using two types of hydrodynamical
simulations: \eagle{}, which resolves a large cosmological volume, and
the \auriga{} suite, which are very high resolution zoom-in
simulations of individual MW-sized haloes. The present-day 11 most
massive satellites are predominantly (75\%) accreted in single events,
14\% in pairs and 6\% in triplets, with higher group multiplicities
being unlikely. Group accretion becomes more common for fainter
satellites, with 60\% of the top 50 satellites accreted singly, 12\%
in pairs, and 28\% in richer groups. A group similar in stellar mass
to the Large Magellanic Cloud (LMC) would bring on average 15 members
with stellar mass larger than $10^4\Msun$. Half of the top 11
satellites are accreted along the two richest filaments. The accretion
of dwarf galaxies is highly anisotropic, taking place preferentially
perpendicular to the halo minor axis, and, within this plane,
preferentially along the halo major axis. The satellite entry points
tend to be aligned with the present-day central galaxy disc and
satellite plane, but to a lesser extent than with the halo
shape. Dwarfs accreted in groups or along the richest filament have
entry points that show an even larger degree of alignment with the
host halo than the full satellite population. We also find that having
most satellites accreted as a single group or along a single filament
is unlikely to explain the MW disc of satellites.
\end{abstract}


\section{Introduction} \label{sect:intro}
One of the fundamental predictions of the standard cosmological model,
$\Lambda$ cold dark matter (\lcdm{}), is that dark matter (DM) haloes
grow hierarchically, from the accretion of many lower mass haloes
\citep[e.g.][]{Ghigna1998,Springel2008}, which, once accreted, are
referred to as substructures or subhaloes. The substructures can
survive and orbit their parent halo for a long time, and ultimately
they will either merge with or be tidally disrupted by their host halo
\citep[e.g.][]{Gao2004,Angulo2009,van_den_Bosch2017}. The MW and
Andromeda (M31) are observed to host around 50 and 40 satellite
galaxies \citep{McConnachie2012}, respectively, the former of which is
a subset only of the expected $\sim$120 satellites after
incompleteness corrections \citep{Newton2017}. These satellite
populations provide a crucial window into hierarchical structure
formation, and phenomena such as tidal stripping, strangulation and
ram pressure stripping \citep{Simpson2017}.

Despite being an area of intense study, there are many questions
related to the infall, orbital evolution and tidal disruption of
satellite galaxies that are poorly understood. Here, we focus on the
former aspect, the accretion of satellite galaxies into MW-mass
haloes, and study the statistics of group and filamentary accretion,
the preferential directions along which satellite accretion takes
place, and the implications for the present-day satellite
distribution.

Accretion of galaxy groups is crucial for understanding the MW
satellite populations, especially at the very faint end of the stellar
mass function where ${\sim}20$ new dwarf galaxies were discovered
recently in the Dark Energy Survey (DES;
\citealt{Bechtol2015,Drlica-Wagner2015,Kim2015a,Kim2015b,Koposov2015,Luque2016}),
the Survey of the MAgellanic Stellar History (SMASH;
\citealt{Martin2015}), Pan-STARRS \citep{Laevens2015}, ATLAS
\citep{Torrealba2016} and MagLitesS \citep{Drlica-Wagner2016}. Many of
these recent discoveries are likely to be associated with the Large and
Small Magellanic Clouds (LMC and SMC, respectively), which themselves
are very likely to have fallen in as a group
\citep{Kallivayalil2013}. However, it is yet unclear how many and
which of the MW satellites fell in with the LMC, which, given its
large total mass, is expected to bring a sizeable population of
satellites. \citet{Jethwa2016} inferred that around half of the DES
satellites fell in with the LMC and that as much as 30\% of all MW
satellites could have been brought by the LMC. However,
\citet{Deason2015} and \citet{Sales2017} predicted that on average
only 7\% and 5\%, respectively, of Galactic satellites were
associated with the LMC at infall, although the exact percentage can
range from 1 to 25\% and it is very sensitive to the poorly
constrained LMC total mass \citep[for a compilation of LMC mass
estimates see][]{Penarrubia2016}.

Satellites that fell in together have correlated orbits, which can
have important implications for the present-day spatial and kinematic
distribution of MW and M31 satellites: both Local Group giant galaxies
have highly anisotropic and flattened satellite distributions,
so-called planes of satellite galaxies
\citep{Kunkel1976,Lynden-Bell1976,Lynden-Bell1982,Kroupa2005,Conn2013,Ibata2013};
many of the Galactic classical satellites have nearly co-planar orbits
\citep{Pawlowski2012a}; and the MW classical dwarfs show a tangential
velocity excess indicative of circularly biased orbits
\citep{Cautun2017}. Group accretion, although uncommon
\citep{Wang2013}, may explain one or more of these observed features
of the MW and M31 satellite populations
\citep[][although \citealt{Metz2009} claim that rich groups of dwarfs are not compact enough to generate a thin plane of satellites]{Li2008,Wang2013,Smith2016}. Similar to group accretion,
correlated satellite orbits can arise from the accretion of multiple
satellites along the same filament of the cosmic web, which is
expected to be a common occurrence
\citep[e.g.][]{Aubert2004,Knebe2004,Libeskind2005,Zentner2005b,Deason2011,Wang2014}. Filamentary
accretion is often thought to be responsible for the MW and M31 plane
of satellite galaxies \citep[][]{Libeskind2005,Buck2015,Cautun2015,Ahmed2017} and
for the surplus of satellites that have co-planar orbital planes
\citep[][although this explanation has been questioned, e.g. see \citealt{Pawlowski2012b}]{Libeskind2009,Lovell2011,Cautun2015b}.

Group infall and accretion along filaments are important for the
preprocessing of dwarf galaxies, especially the very faint
ones. Around half of the MW and M31 faint dwarfs could have been
accreted by another low-mass group before final infall into MW/M31,
and thus could have been subject to star formation quenching and tidal
disruption before being accreted into their MW-mass host halo
\citep{Wang2013,Wetzel2015,Wheeler2015}. Group infall can also enhance the chance
of satellite-satellite mergers of MW-mass haloes, with most such
mergers taking place shortly after accretion \citep{Deason2014}. Also,
accretion on to filaments before the further infall into the MW/M31
halo can lead to gas stripping and star formation quenching of faint
dwarfs \citep{Benitez-Llambay2013,Simpson2017}.

In this paper, we study the prevalence of group and filamentary
satellite accretion of MW-mass haloes. We use the \eagle{}
\citep{Schaye2015,Crain2015} hydrodynamical cosmological simulations
which, because of its large volume, has a large sample of MW-mass
haloes with luminous satellite population similar to the Galactic
classical satellites. For studying even fainter dwarfs, we use the
\auriga{} \citep{Grand2017} suite of zoom-in hydrodynamic
resimulations of 30 MW-mass haloes, which allows us to study the
orbital history of dwarfs with stellar masses as low as
${\sim}10^5\Msun$ (for the main \auriga{} sample) and
${\sim}10^4\Msun$ (for a subset of six \auriga{} haloes resimulated at
even higher resolution). We also quantify the anisotropy of satellite
accretion, and how these anisotropies are connected to group and
filamentary infall. We end with a statistical analysis of the impact
of group and filamentary accretion on the flattening of the MW
classical satellite distribution and on the extent to which it
increases the number of satellites with highly clustered orbital
poles.

The paper is organized as follows. Section~\ref{sect:simul} reviews
the simulations used in this work and describes our sample selection;
Section~\ref{sect:result} presents our main results; we conclude with
a short summary and discussion in Section~\ref{sect:conclusion}.

\vspace{-.3cm}
\section{Simulation and methods}
\label{sect:simul}
We make use of two sets of simulations: \eagle{} and \auriga{}.
\eagle{} is the main cosmological hydrodynamical simulation (labelled
\simRef{}) performed as part of the \eagle{} project
\citep{Schaye2015, Crain2015}; it consists of a periodic cube of
$100\Mpc{}$ side length and follows the evolution of $1504^3$ DM
particles and an initially equal number of baryonic particles. The DM
particles have a mass of $9.7\times 10^6 \Msun$, and the gas particles
have an initial mass of $1.8\times 10^6 \Msun$. The simulation uses
the \textit{Planck} cosmology \citep{Planck2014} with cosmological
parameters: $\Omega_{\rm m}=0.307, \Omega_{\rm b}=0.04825,
\Omega_\Lambda=0.693,h=0.6777,\sigma_8=0.8288$ and $n_{\rm s}=0.9611$.
The \eagle{} simulation was performed using a modified version of the
\textsc{gadget} code \citep{Springel2005c}, which includes
state-of-the-art smoothed particle hydrodynamics methods
\citep{DallaVecchia2012,Hopkins2013,Schaller2015a}. The main physical
processes implemented in \eagle{} were calibrated to reproduce the
present-day stellar mass function and galaxy sizes, as well as the
relation between galaxy stellar masses and supermassive black hole
masses \citep{Crain2015, Schaye2015}. See \citet{Schaye2015} for a
more detailed description of the baryonic processes implemented in
\eagle{}.

\auriga{} is a suite of zoom-in hydrodynamical cosmological
simulations of isolated MW-mass haloes \citep{Grand2017} within the
\textit{Planck} cosmology. The suite consists of 30 medium-resolution
simulations, which we refer to as \auriga{} level-4, that have an
initial gas particle mass of $5\times 10^4 \Msun$ and a DM particle
mass of $3\times 10^5 \Msun$. Six of these haloes, which we refer to
as \auriga{} level-3, have been resimulated with an eight times higher mass
resolution. The simulations were performed using the {\it N}-body,
magnetohydrodynamics code \textsc{arepo} \citep{Springel2010}, and
included many physical processes relevant for galaxy formation such as
black hole accretion and feedback, stellar and chemical evolution,
stellar feedback, metallicity-dependent cooling, star formation and
magnetic fields. The properties of \auriga{} galaxies show a good
agreement with observational data: they have flat rotation curves;
realistic present-day star formation rates and reproduce the
mass-metallicity relation (see \citealt{Grand2017} for a more detailed
comparison).

In both simulations, haloes were identified using the
friends-of-friends (FOF) algorithm \citep{Davis1985} with a linking
length of $0.2$ times the mean particle separation. The haloes were
further processed to identify gravitationally bound substructures,
which was performed by applying the \textsc{subfind} code
\citep{Springel2001,Dolag2009} to the full matter distribution (DM,
gas and stars) associated with each FOF halo. The resulting population
of objects was split into main haloes and subhaloes. The main haloes
correspond to the FOF substructure that contains the particle with
the lowest gravitational energy, and its stellar distribution is
classified as the central galaxy. The main haloes are characterized in
terms of the mass, $M_{200}$, and the radius, $R_{200}$, corresponding
to an enclosed spherical overdensity of $200$ times the critical
density. The remaining subhaloes are classified as satellite
galaxies. The position of each galaxy, for both centrals and
satellites, is given by the particle with the lowest gravitational
potential energy.

To trace the evolution of galaxies across multiple simulation outputs,
we used the \eagle{} and \auriga{} galaxy merger trees
\citep{Springel2005a,De_Lucia2007,McAlpine2016,Qu2017}. Galaxies form and evolve within their
host haloes, so tracing them across snapshots is analogous to tracing
the evolution of their host haloes. The \eagle{} merger trees were
constructed by applying the \textsc{d-trees} algorithm
\citep{Jiang2014} to \textsc{subfind} subhalo catalogues across all
simulation snapshots. This consists of uniquely linking a subhalo with
its descendant across two consecutive simulation outputs. A subhalo
descendant, and hence that of the galaxy residing in that subhalo, is
identified by tracing where the majority of the most bound particles
are located in the successive snapshot. While each subhalo has a
unique descendant, it can have multiple progenitors. To trace back the
temporal evolution of a $z=0$ galaxy, we follow the \textit{main
progenitor} branch, which for any snapshot is defined as the branch
with the largest total mass summed across all the earlier snapshots.

\subsection{Sample selection}
\begin{figure}
 \plotone{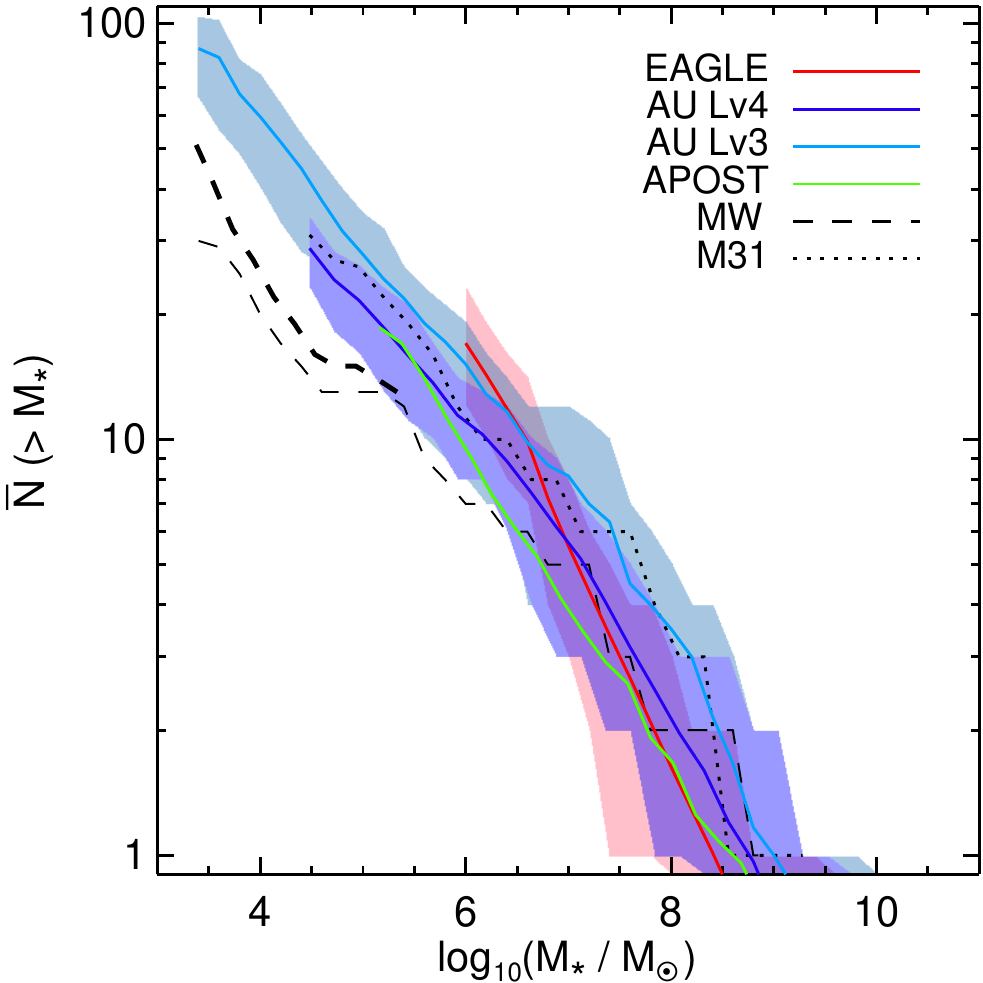}
 \caption{The satellite stellar mass function of the MW-mass haloes
studied here. It shows the average number of luminous satellites per
host within $300\kpc$ as a function of satellite stellar mass,
$M_\star$. The solid lines show median estimates and the shaded
regions show 16th to 84th percentiles for the \eagle{} (red),
\auriga{} level-4 (purple), and \auriga{} level-3 (blue)
simulations. We also indicate the median satellite stellar mass
function of \apo{} \citep[solid green][]{Sawala2016}. The black dotted
and dashed lines give the observed satellite stellar mass function
within $300\kpc$ of the MW and M31, respectively. The thicker dashed
line illustrates the incompleteness-corrected MW satellite stellar
mass function \citep{Newton2017}.
 }
 \label{fig:mf-all}
\end{figure}

To identify systems similar to the MW and M31, we start by selecting
in the \eagle{} simulation haloes with mass, $M_{200} \in [0.3, 3]
\times 10^{12}\Msun$. The wide mass range is motivated by the large
uncertainties in the total mass of the MW \citep[e.g.][]{Fardal2013,
Cautun2014a, Piffl2014, Wang2015, Han2016} and the need to have a
large sample of such systems. We further select isolated haloes by
excluding any central galaxy that has a neighbour within $600\kpc$
with a stellar mass larger than half their mass. We also restrict our
selection to haloes that, like the MW, have at least 11 luminous
satellites within a distance of $300 \kpc$ from their central
galaxy. \eagle{} contains 1080 host haloes that satisfy all the
selection criteria; the sample has a median halo mass, $M_{200} \sim
1.2 \times 10^{12}\Msun$, and, on average, $15$ luminous satellites
per halo. For the \auriga{} simulation, we use all the 30 systems,
which were selected in the first place to be isolated and to have halo
masses similar to the MW halo mass. For both simulations, we only
consider luminous satellites, defined to be subhaloes with at least
one star particle.
Selecting substructures with one or more star particles means
that these substructures are luminous and that we capture any biases 
between luminous and dark subhaloes, if such biases are present.
Apart from random effects arising from the stochastic
nature of star formation in the simulations, this sample selection
is robust. A higher resolution simulation with identical subgrid
physics would, on average, assign the same luminosity to the same haloes in
the current simulation although with more (less massive) star
particles.

\reffig{fig:mf-all} investigates the satellite stellar mass function
of the two simulations within a distance of $300\kpc$ from each
central galaxy. We find good agreement between \eagle{} and \auriga{}
medium resolution, with the only noticeable discrepancy being for
$M_\star<10^{7}\Msun$, which is close to the resolution limit of
\eagle{}. The stellar mass function of the \auriga{} high-resolution
sample is systematically higher than both \eagle{} and \auriga{}
level-4 ones; this likely is due to the small number (6) of
high-resolution system and due to the fact that these systems have halo
masses, on average, 10\% more massive than those in the full \auriga{}
sample. The \eagle{} and \auriga{} results are consistent with the
\apo{} ones \citep{Sawala2016}, which is another suite of zoom-in
simulations of paired MW-mass haloes chosen to resemble the Local
Group \citep{Fattahi2016}.

\reffig{fig:mf-all} also shows that the dwarf stellar mass functions
found in our simulations are consistent with the ones observed around
the MW and M31 \citep[see also][]{Sawala2016,Simpson2017}, which we
take from the \citet{McConnachie2012} compilation. The agreement is
especially good with the M31 observations, whereas the MW data is
systematically lower, especially for satellites less massive than
$10^{7}\Msun$. The MW satellite stellar mass function is affected by
incomplete sky coverage; accounting for this using the
\citet{Newton2017} predictions pushes up the faint end of the MW dwarf
count, but not enough to fully account for the difference. The
discrepancy could be due to combination of factors, such as the total
MW halo mass being lower than that of our sample, the MW having an
atypically low number of satellites for its mass, or a higher than
accounted for observational incompleteness of MW surveys as argued by
\citet{Yniguez2014}.

\label{sec:multi}
\begin{figure*}
 \vspace{-2.cm}
 \plotthree{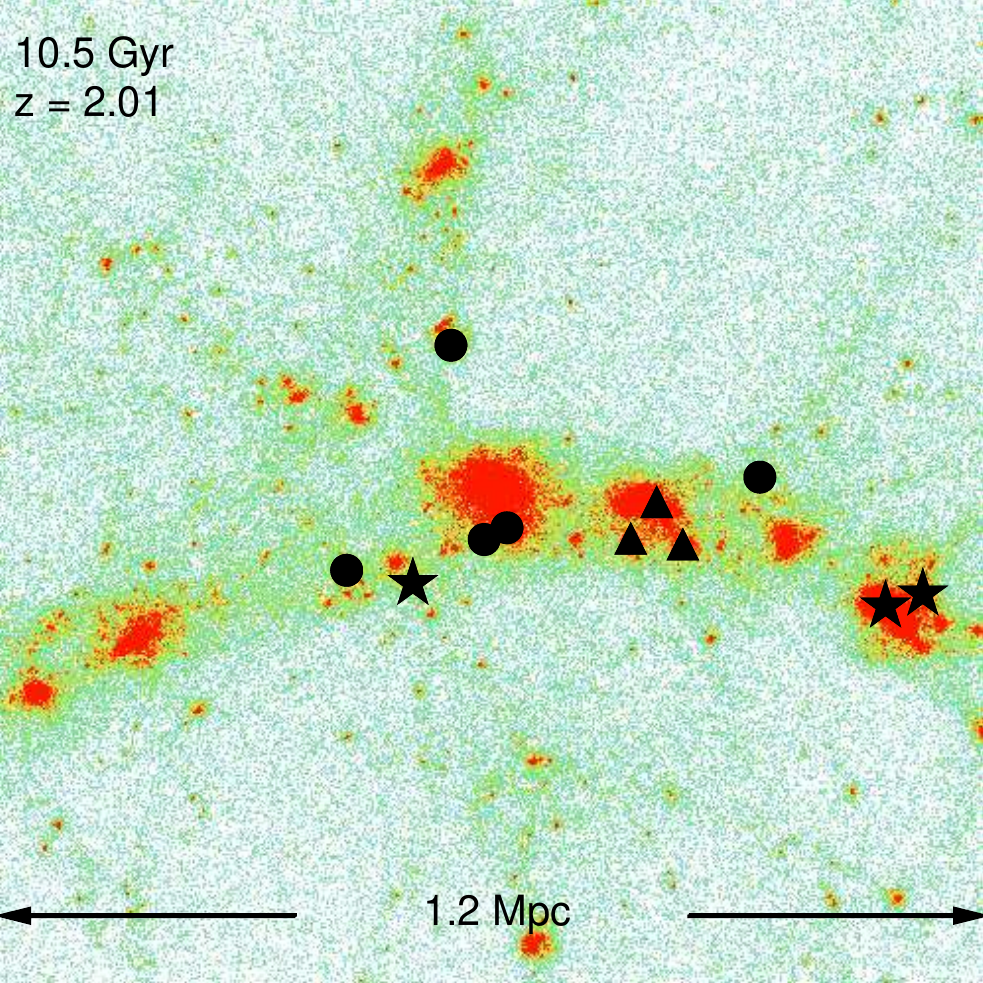}{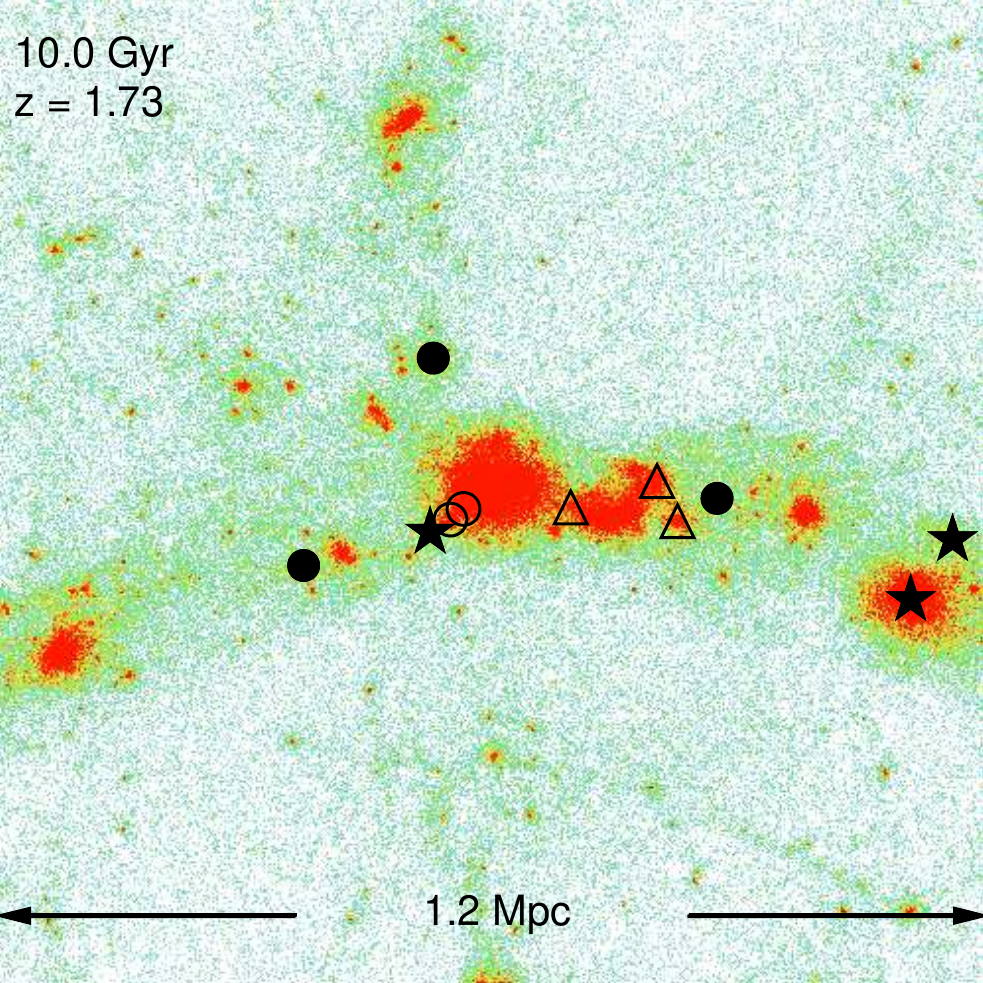}{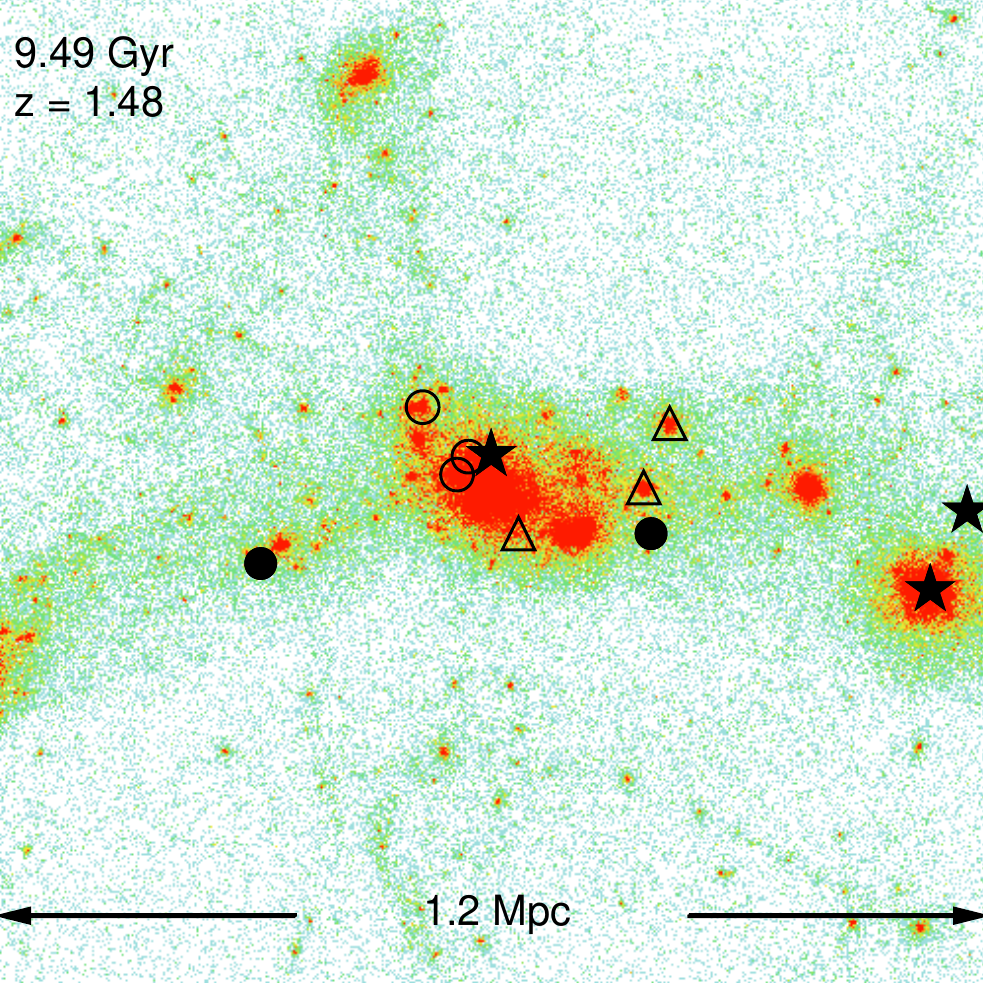}\vspace{0.05cm}
 \plotthree{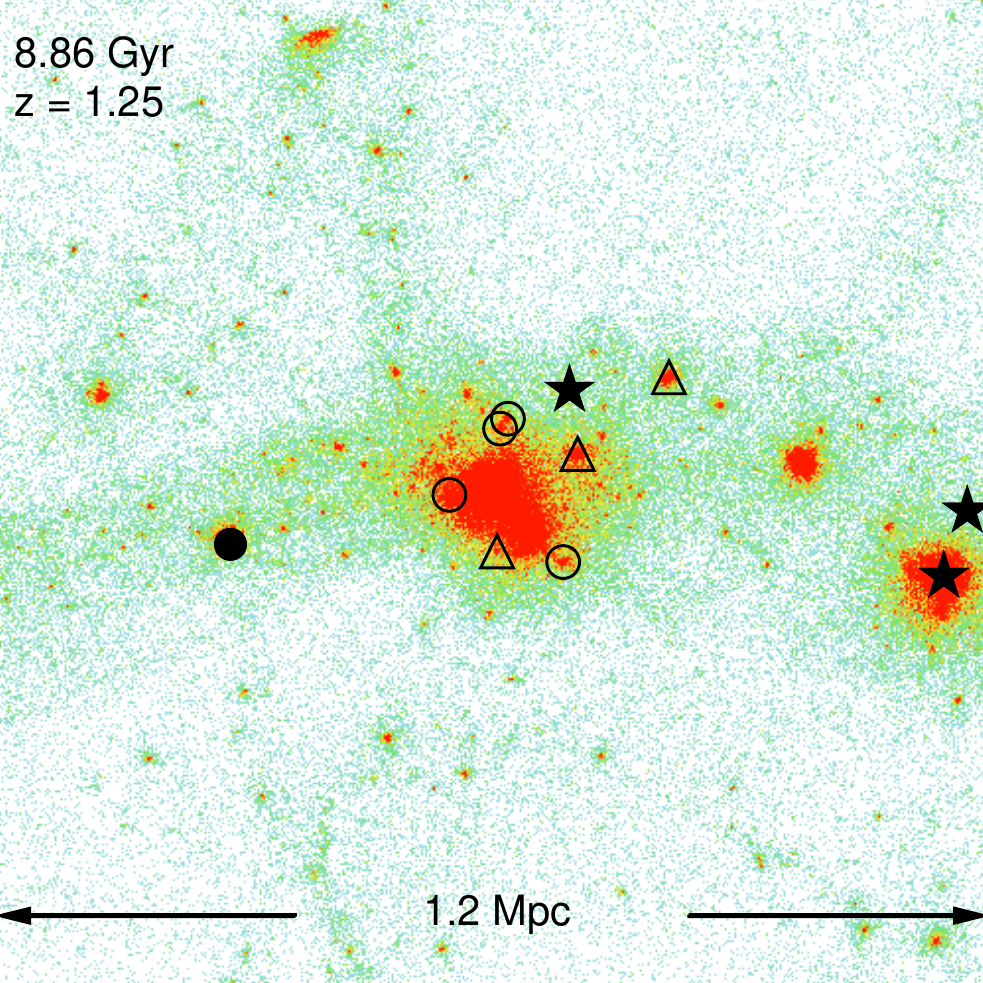}{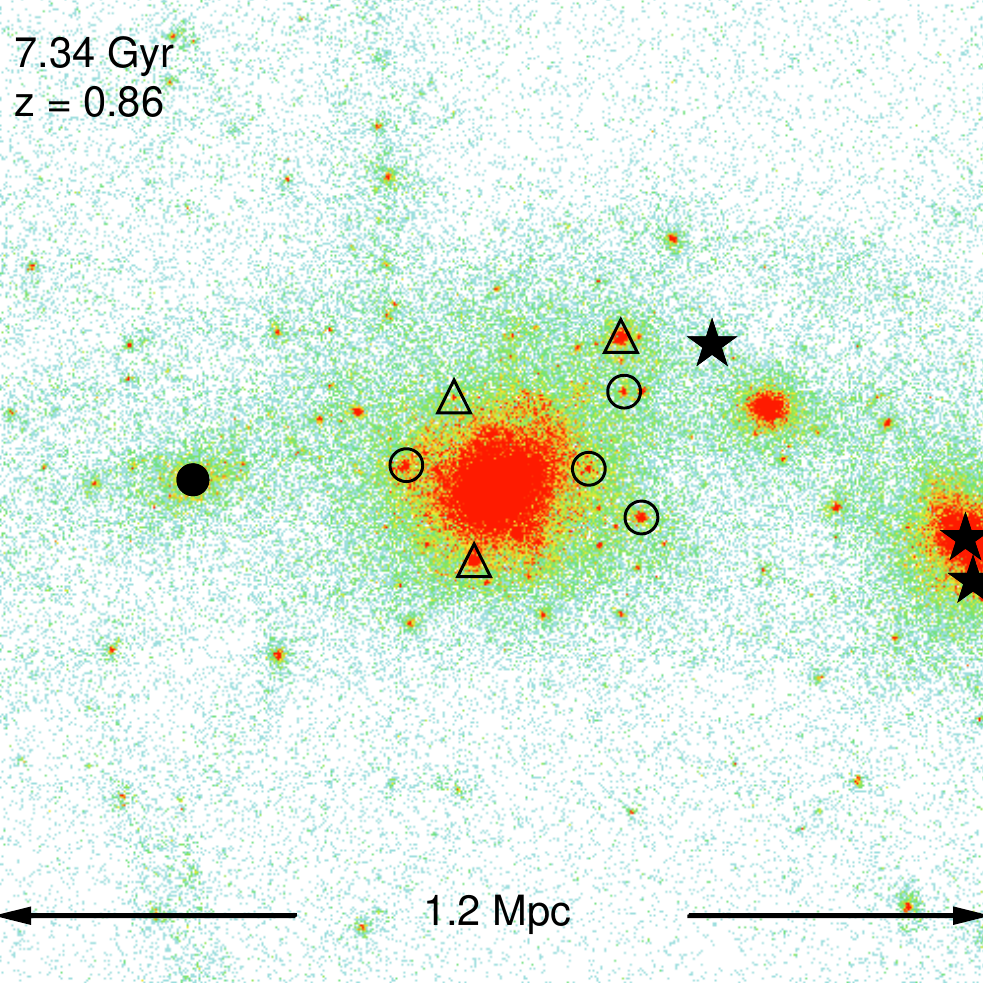}{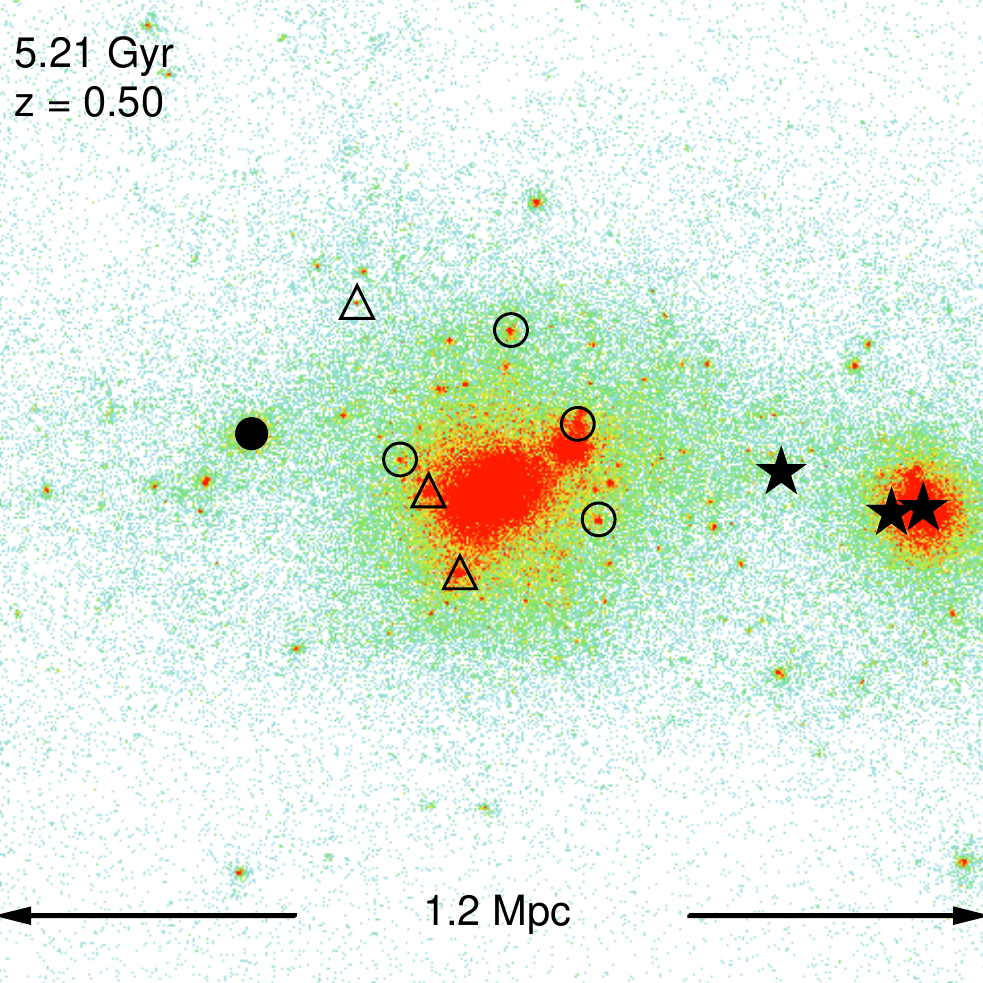}\vspace{0.05cm}
 \plotthree{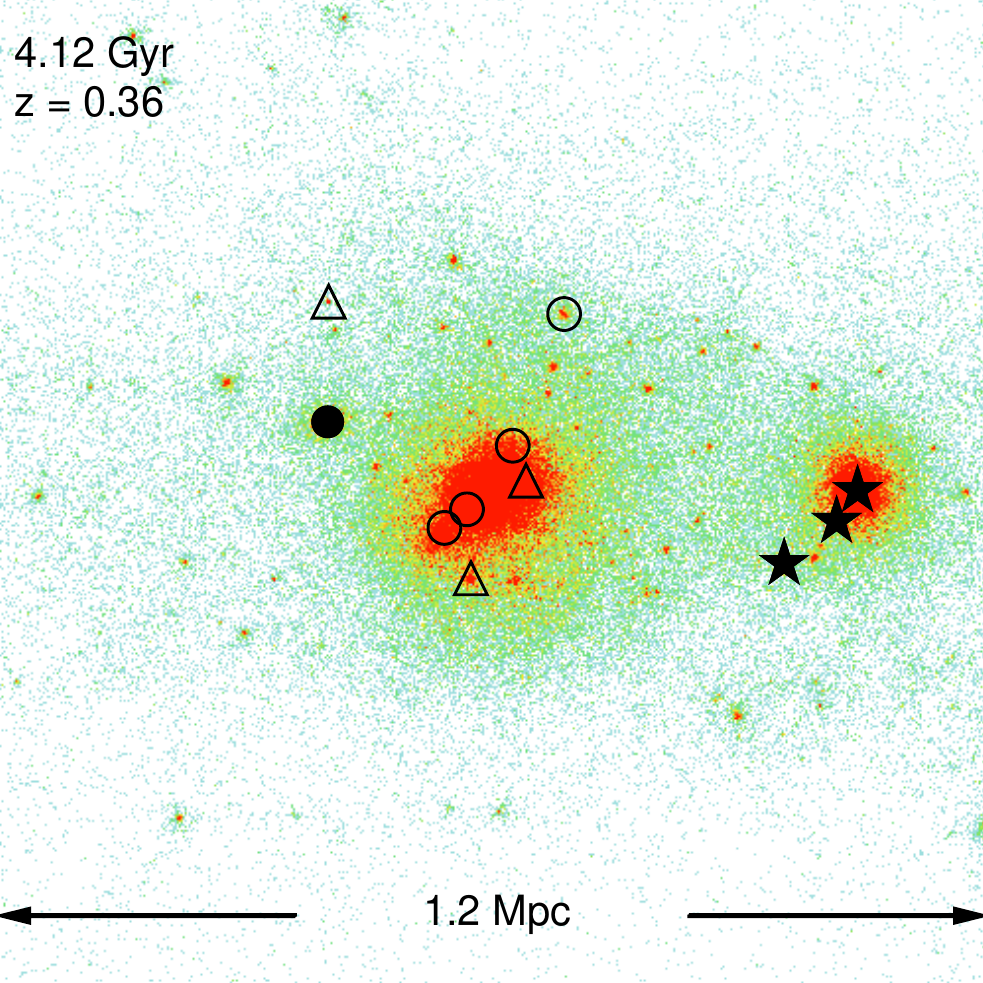}{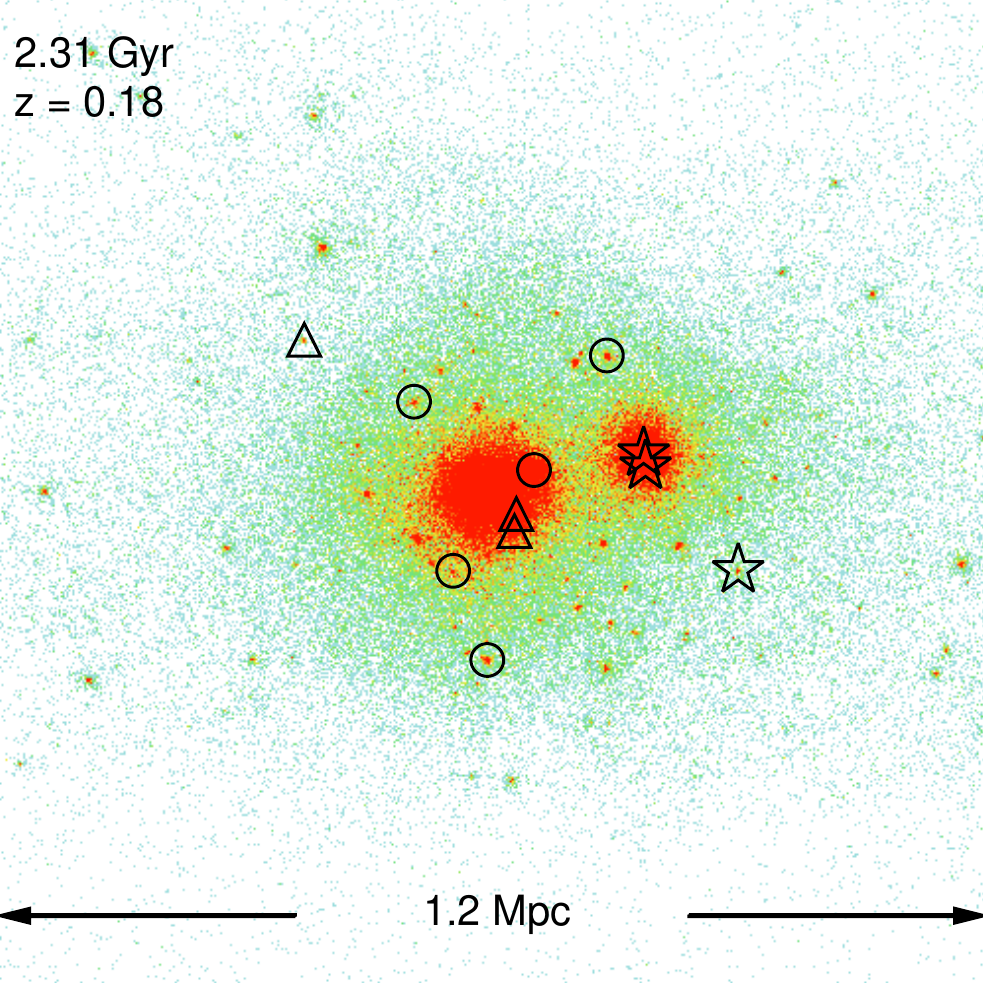}{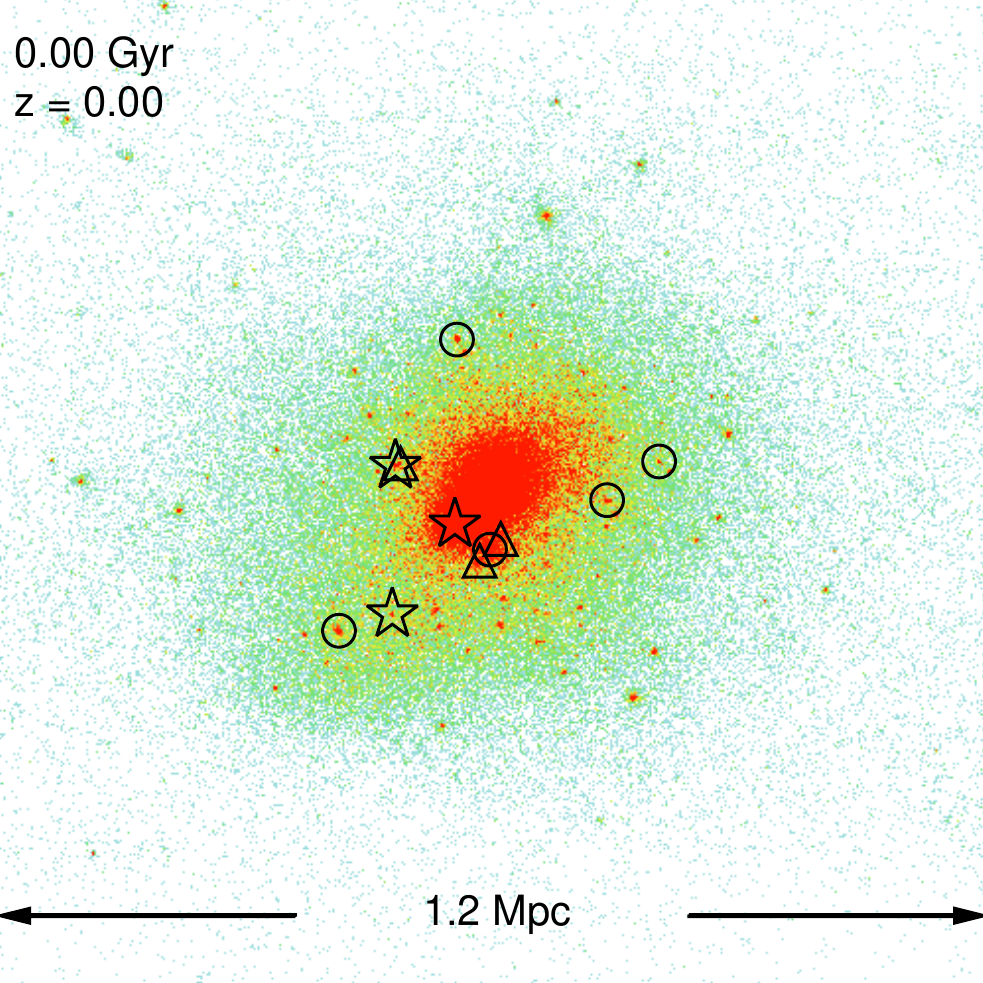}
 \caption{Evolution of a \eagle{} MW-mass system since $z=2.01$ to
present day. The colours indicate the DM distribution within a
$1.2~\rm{Mpc}$ physical box centred on the halo position, with red
showing high-density regions. The black symbols indicate the positions
of the progenitors for the 11 $z=0$ satellites with the largest
stellar mass. These 11 satellites were accreted in two groups of
multiplicity, $\rm m_{11} = 3$, with the members of those groups shown
as star and triangle symbols, while the remaining satellites were
accreted singly and are shown as circles. The satellite
progenitors are shown as filled symbols before infall, and as open
symbols after infall into the host halo.
 }
 \label{fig:sketch}
\end{figure*}

\section{Results}
\label{sect:result}
Here we study the fraction of satellites that were accreted as groups
or along the same filament, after which, we quantify the anisotropies
in the accretion of satellites by investigating the alignment of the
infall direction of each satellite with the preferential axes of its
host systems. We end with an analysis of the connection between group
and filamentary accretion with the MW disc of satellite galaxies,
i.e. the structures present in the spatial and kinematic distribution
of the Galactic satellites.

\subsection{Multiplicity of satellite accretion}

\begin{figure}
 \plotone{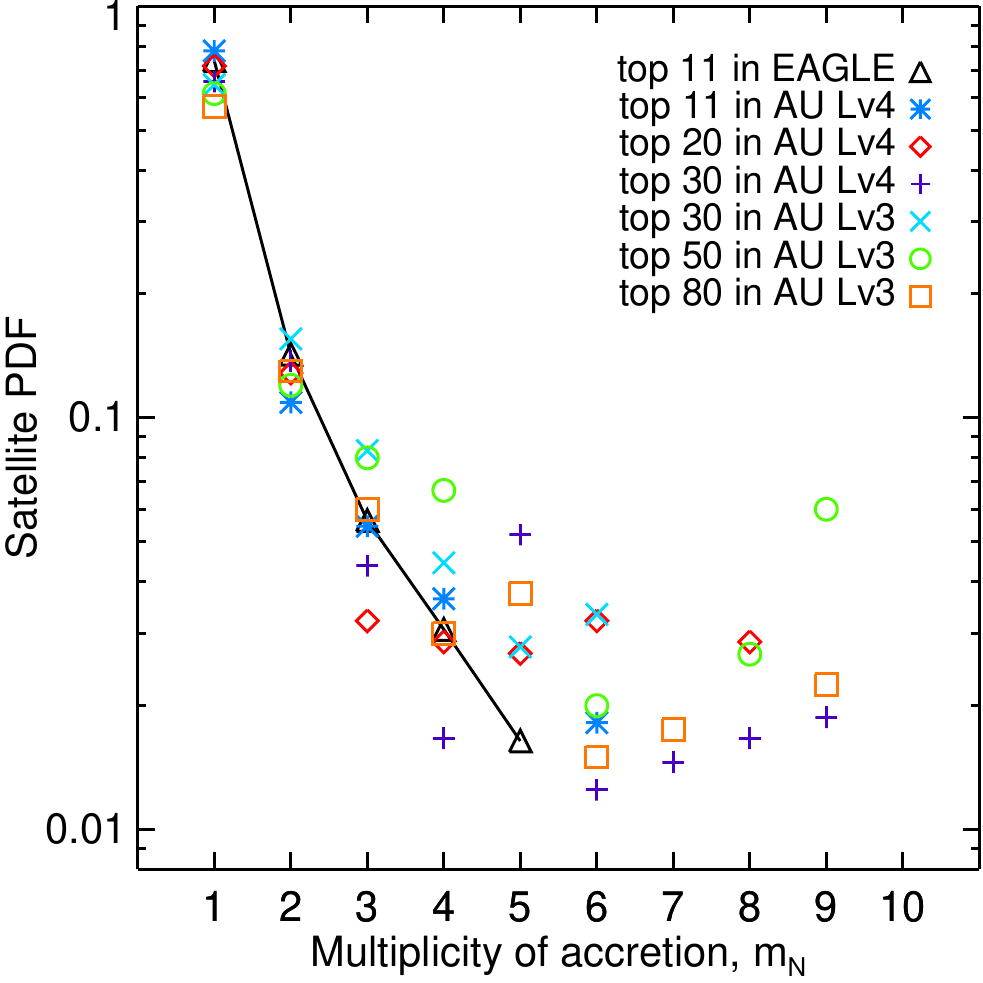} \\[.2cm]
 \plotone{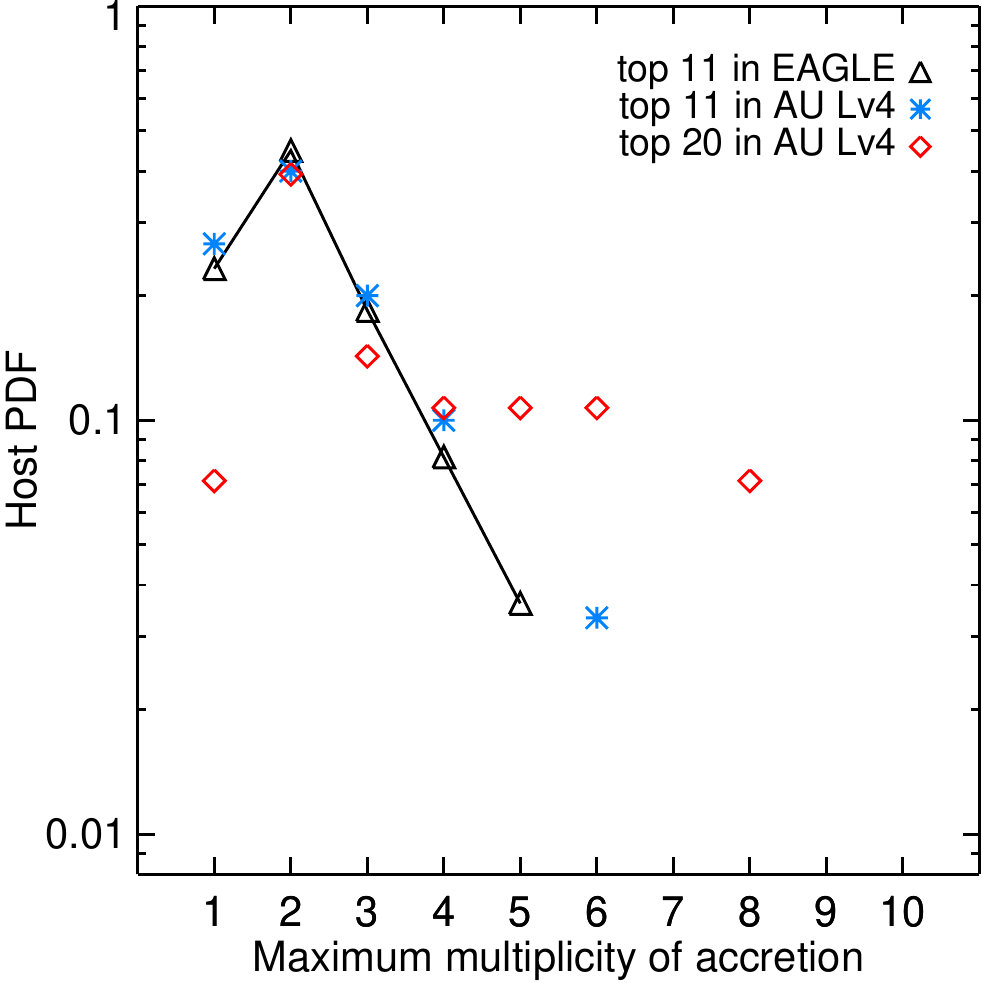}
 \caption{ The PDF of the multiplicity of accretion for the top $N$
satellites with the largest $z=0$ stellar mass. Top panel: the
vertical axis gives the fraction of satellites accreted in groups of
that given multiplicity, with $m_{N}=1$ corresponding to singly
accreted satellites. The various symbols correspond to different
values of $N$ and to different simulations, which are \eagle{}, and
\auriga{} medium (level-4) and high resolution (level-3), as indicated
in the legend. For readability, the \eagle{} results are shown as
triangles connected with a solid line. Bottom panel: the
vertical axis gives the fraction of hosts as a function of their
maximum multiplicity of accretion, i.e. the number of objects in the
richest accreted satellite group. We show results only for \eagle{}
(1080 hosts) and \auriga{} level-4 (30 hosts) using the same symbols
as in the top panel.
 }
 \label{fig:frac}
\end{figure}

\begin{figure}
 \plotone{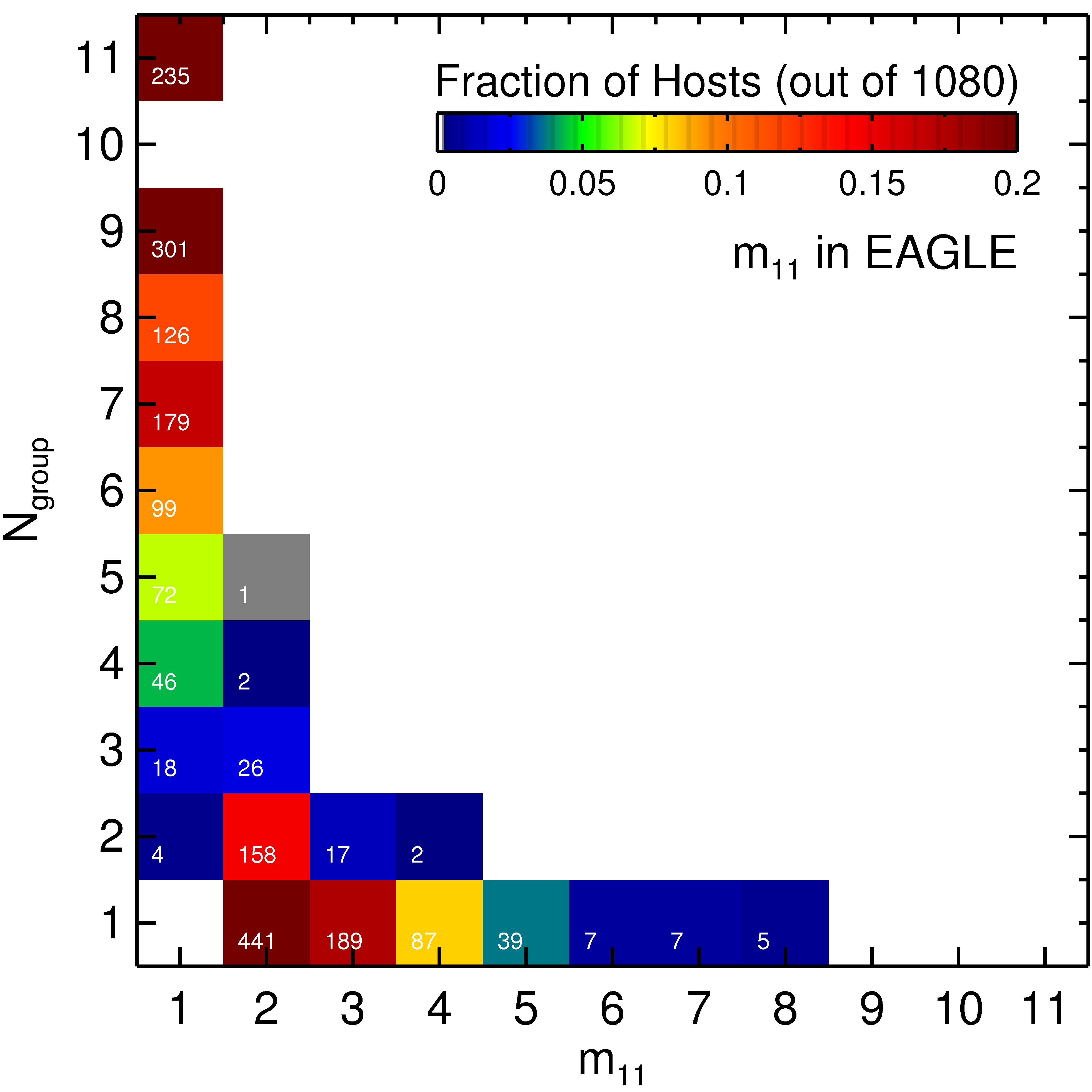} \\[.2cm] \plotone{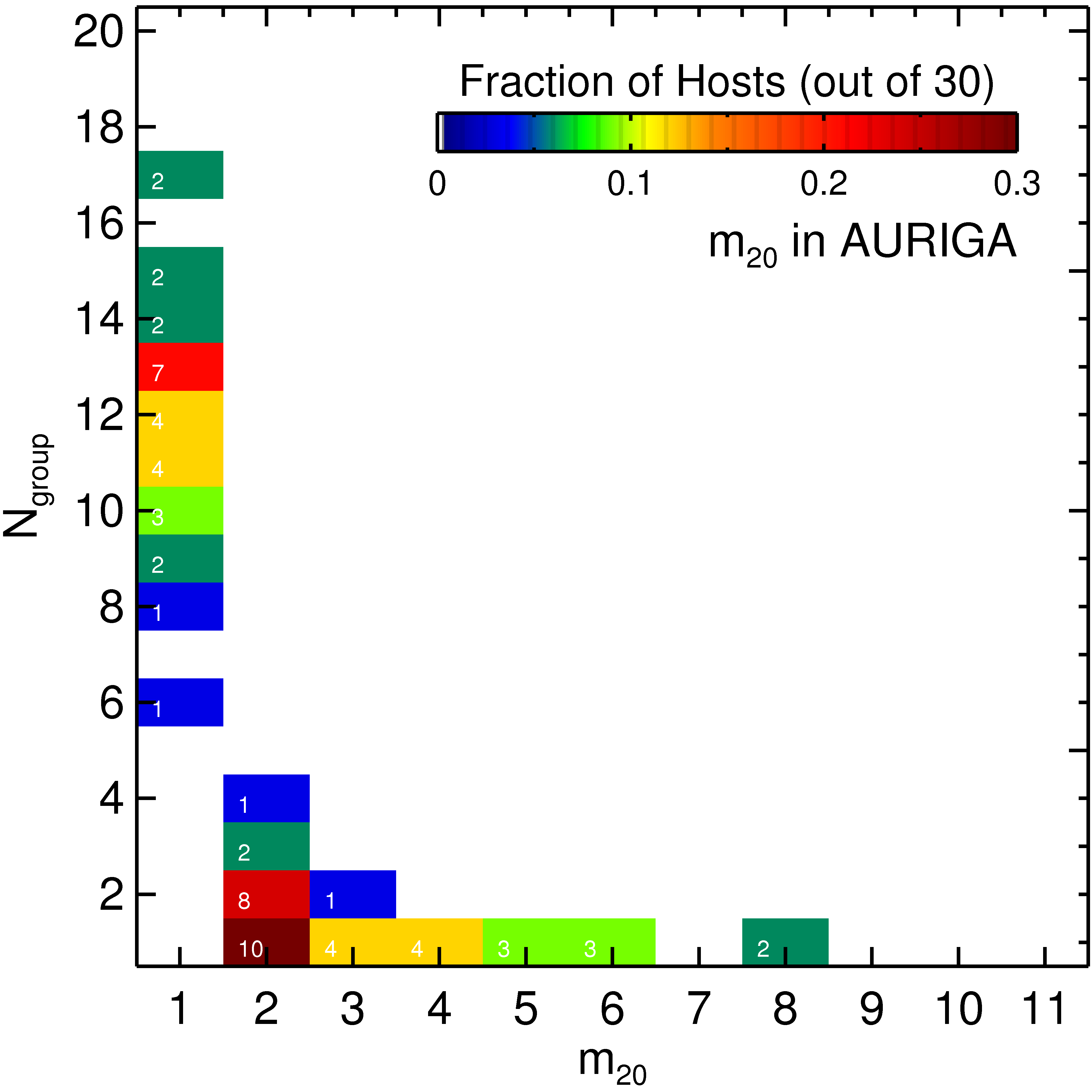}
 \caption{ Histogram of the multiplicity of satellite accretion in the
\eagle{} (top panel) and the \auriga{} level-4 (bottom panel) MW-mass
halo samples. The top and bottom panels show the multiplicity of
accretion, $m_{11}$ and $m_{20}$, for the respectively 11 and 20
satellites with the largest present-day stellar mass. The colours
indicate the fraction of haloes with a given satellite accretion
history, e.g. in the top panel the $(m_{11}=1,~N_{\rm {group}}=6)$
point shows that 9\% of haloes had accreted six groups with
multiplicity 1, and the $(m_{11}=2,~N_{\rm {group}}=2)$ point shows
that 15\% of haloes accreted two groups of multiplicity 2. The number
inside each histogram entry gives the count of host haloes (out of
1080 for the top panel and out of 30 for the bottom one) with that
satellite accretion history.
 }
 \label{fig:top}
\end{figure}

\begin{figure}
 \plotone{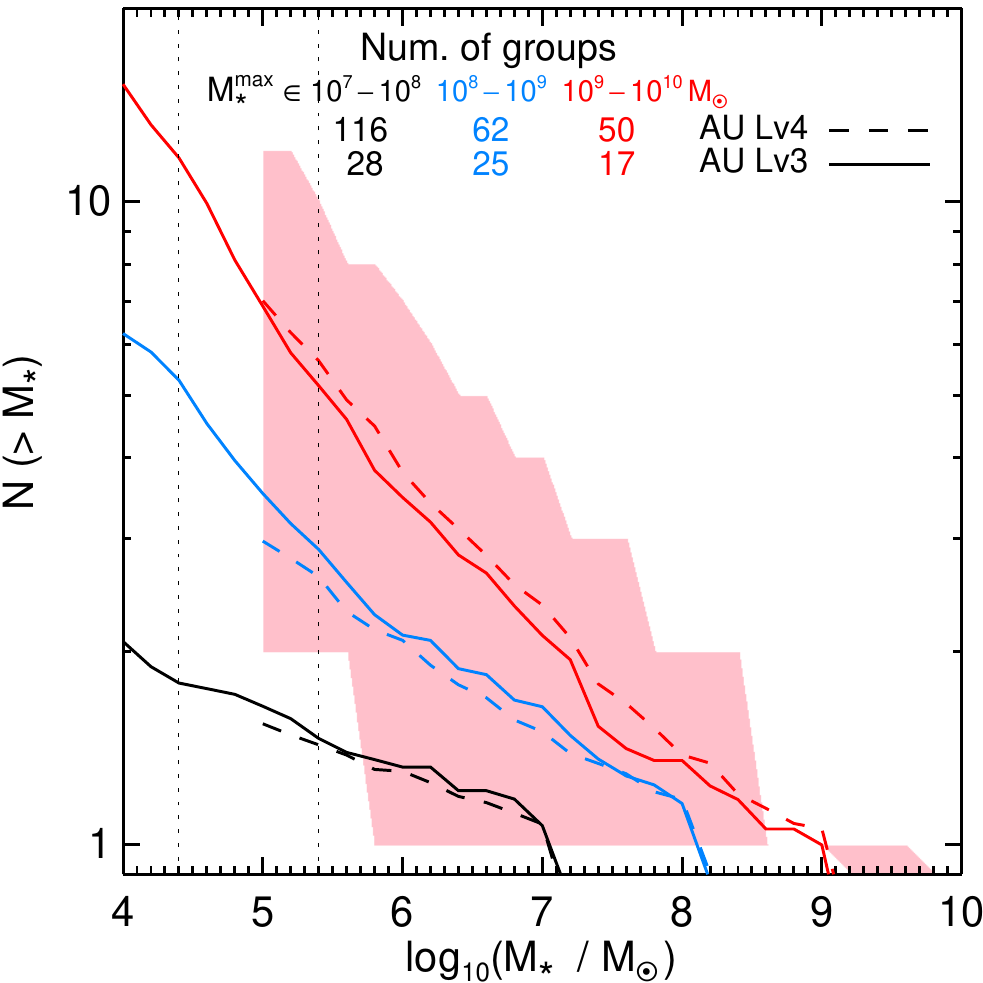}
 \caption{ Stellar mass function of satellites of satellites that were
accreted into MW-mass haloes. The curves are split according to the
stellar mass of the primary object, which is the galaxy in the group
with the largest stellar mass before accretion into the MW-sized host
halo. We show three bins in primary stellar mass, $10^9-10^{10}$
(red), $10^8-10^9$ (blue), and $10^7-10^8$ $\Msun$ (black), for both
\auriga{} level-4 and level-3 simulations, and we give in the legend
the number of groups contained in each subsample. The dashed and solid
lines show the average satellites-of-satellites count for the medium
and the high-resolution simulations. The shaded region shows the 16th
to 84th percentile range, which, for clarity, we only show for the
most massive subsample in \auriga{} level-4. The vertical dotted lines
indicate five times the initial gas element mass for the two resolution
levels.
 }
 \label{fig:mf-acc}
\end{figure}

Our goal is to study the accretion of the present-day brightest $N$
satellites, which we select as the $N$ satellites with the largest
$z=0$ stellar mass and that are within a distance of $300\kpc$ from
the central galaxy. We refer to these objects as the top $N$
satellites and we vary $N$ from 11, corresponding to the MW classical
satellites, to 80, which is determined by the smallest number of
satellites across each of the six \auriga{} high-resolution
haloes. For each of the top $N$ satellites, we calculate the
multiplicity of accretion, $m_N$, as the number of top $N$ satellites
that were part of the same group at accretion into the host halo.

For each central and satellite galaxy, we trace their formation
history using the \eagle{} and \auriga{} merger trees for the most
massive progenitor. Starting at high redshift, we follow forward the
merger trees of each satellite in tandem with the merger tree of its
central galaxy, until we find the first snapshot where the satellite
and the central are part of the same FOF group; this corresponds to
the snapshot when the satellite was first accreted on to its $z=0$ host
halo.\footnote{In a small number of cases satellite galaxies may drift
in and out of the host FOF halo. Even in those cases, we define the
accretion time as the first time the satellite enters the $z=0$ host
halo.} Then, the group ID in which that satellite was accreted is
given by the FOF halo ID of its progenitor at the snapshot just before
first accretion. We repeat this procedure for the top $N$ satellites
of each host, and, once finished, count how many satellites have the
same group ID just before accretion.
Groups are defined as the subset of galaxies that in the 
snapshot just before accretion were part of the same FOF group. 
We do not require that they be gravitationally bound, so a small 
fraction of groups could potentially 
contain unbound members that fell into the host halo
at the same time and along a similar direction.

\reffig{fig:sketch} illustrates the evolution of the progenitors for
the top 11 present-day satellites of one \eagle{} halo. This system
has two group accretion events each with three satellites that are
part of the top 11 $z=0$ satellites. The satellites accreted in the
two groups are shown as triangle and star symbols, while the remaining
satellites, which were accreted singly, are shown as circles. The
triplet shown with triangles is accreted early, being already part of
the same FOF halo in the second panel of \reffig{fig:sketch}. This
group was probably loosely bound, because, in the subsequent frames,
its three members are spread over most of the halo; however, in the
last two frames, two of these members form a tightly bound pair. The
second triplet, which is shown as stars, has a different evolution
history. Two of its members were a long-lived group since at least the
top-left panel, while the third member falls in along a different
filament and only becomes part of the triplet shortly before
accretion, which takes places at $z\simeq0.2$.

The top panel of \reffig{fig:frac} quantifies the probability
distribution function (PDF) that a satellite was accreted singly,
i.e. $m_N=1$, or as part of a group, i.e. $m_N \geqslant 2$, for
different populations of top $N$ satellites, with $N$ ranging from 11
(corresponding to the classical MW satellites) to 80. The top 11
satellites are predominantly accreted by themselves, which happens in
75\% of cases. Group accretion is dominated by pairs, 14\% of the
time, and triplets, 6\% of the time, whereas rich groups with
$m_N\geq6$ represent 1\% of cases. As we increase $N$ and we study a
larger number of top satellites, we find that a larger fraction of
satellites are accreted in groups. For example, using the high-resolution \auriga{} simulations we find that the top 50 satellites
were accreted singly 60\% of the time, in pairs 12\% of the time,
and in groups of six or more members 12\% of the time. The results
for top 50 satellites, while limited by the small number of systems (6
hosts with 300 satellites), confirm the trend of group accretion to
become more important for the faint satellites and agree with the
trends found by previous studies, based on dissipationless simulations
coupled with semi-analytic galaxy formation models or abundance
matching \citep{Wang2013,Wetzel2015}.

The multiplicity of group accretion can be quantified from the
perspective of the host halo. The bottom panel of \reffig{fig:frac}
shows the fraction of MW-sized hosts that have a given maximum
multiplicity of accretion. When considering the top 11 satellites in
\eagle{}, the most likely outcome is the accretion of one or more
satellite pairs (45\% of cases), followed by hosts that accreted all
their satellites singly (23\% of cases). The accretion probability
of triplets and richer groups decreases rapidly, with 18, 8, 4,
2\% of MW-sized haloes accreting groups with a maximum multiplicity
of 3, 4, 5, and 6 or higher, respectively. As expected, when
considering fainter satellites, such as the top 20, we find that the
multiplicity of the richest group increases.

\reffig{fig:top} presents a detailed histogram of the number of groups
of different multiplicities that were accreted by each host
halo. Focusing first on the top 11 satellites in the \eagle{}
simulation (top panel), we find that 235 hosts (22\%) accreted all
their satellites singly, while less than 13\% (sum of the boxes with
$m_{11}=1$ and $N_{\rm group}\leq5$) of hosts have five or fewer
satellites accreted singly. For pair accretion, i.e. $m_{11}=2$, we
find one extraordinary host that has accreted as many as five pairs, 26
hosts have accreted three pairs, while $\gtrsim15\%$ of the hosts have
accreted two pairs of satellites. Furthermore, by summing the numbers in
the $m_{11}=2$ column, we find that $\approx60\%$ of hosts have
accreted at least one pair of satellites. For multiplicity greater
than 2, we find that almost 20\% and 14\% of hosts have accreted
one group with multiplicity, $m_{11}=3$ and $m_{11}\geq4$,
respectively. The plot suggests that the probability of most of the
top 11 satellites to be accreted as a single group is very small, with
less than 2\% of our MW-mass sample having accreted a group with
multiplicity of 6 or higher.

The bottom panel of \reffig{fig:top} shows the multiplicity of
accretion histogram for the top 20 satellites in the \auriga{} level-4
simulations. Of the 30 \auriga{} systems, two are dominated by singly
accreted satellites, that is the ones with $m_{20} = 1$ and $N_{\rm
group}= 17$, and none of the \auriga{} hosts has only singly accreted
satellites. For multiplicity $\rm m_{20} = 2$, we find that
$\approx70\%$ of the haloes (21 out of 30) have accreted at least one
pair of satellites, which is 10\% higher than the fraction for the
top 11 satellites. For higher multiplicities, the small number of
\auriga{} hosts limits the extent to which we can make statistically
robust assertions.

We have shown that a sizeable fraction of the top $N$ satellites were
accreted in groups. The richness of such groups is likely correlated
to the total halo mass of the group, with more massive groups bringing
in a larger number of satellite galaxies. This raises the question:
how many satellites would an LMC or an SMC mass galaxy bring with it? We
investigate this in \reffig{fig:mf-acc}, where we plot the satellite
stellar mass function of groups at infall. The satellites of these
groups, once accreted, become satellites of satellites. We further
split the groups into subsamples according to the stellar mass of the
most massive member of the group, which can also be done in
observations. Groups in which the dominant galaxy has a stellar mass
of $10^9-10^{10}\Msun$, which includes the LMC, bring in a
considerable contingent of satellites, and have $3$, $7$
and $15$ members with stellar masses higher than respectively $10^6$, 
$10^5$ and $10^4\Msun$, which is in agreement with the abundance
matching predictions of \citet{Dooley2017}. 
We checked that restricting the selection criteria to groups
where the dominant galaxy has a stellar mass in the 
range $1-4 \times 10^9\Msun$, which corresponds to the LMC stellar mass,
we get the same satellites-of-satellites mass function.
Compared to the MW, which
for the same stellar masses has about 7, 13 and 20 satellites, the LMC
could have brought a modest, but non-negligible, number of its own
satellites. The scatter in the satellite stellar mass function of
LMC-mass groups is considerable, which is probably a manifestation of
the large scatter between stellar mass and halo mass for LMC-sized
dwarf galaxies \citep{Schaye2015,Sawala2015}, with the satellite
luminosity function expected to correlate more strongly with the total
halo mass. Groups that host less massive dominant galaxies bring in
fewer satellites, with SMC-sized (stellar mass range of
$10^8\sim10^9\Msun$) and Fornax-sized (stellar mass range of
$10^7\sim10^8\Msun$) groups bringing respectively six and two members more
massive than $10^4\Msun$.

\subsection{Filamentary accretion}
\label{sec:filaments}
\begin{figure}
 \plotone{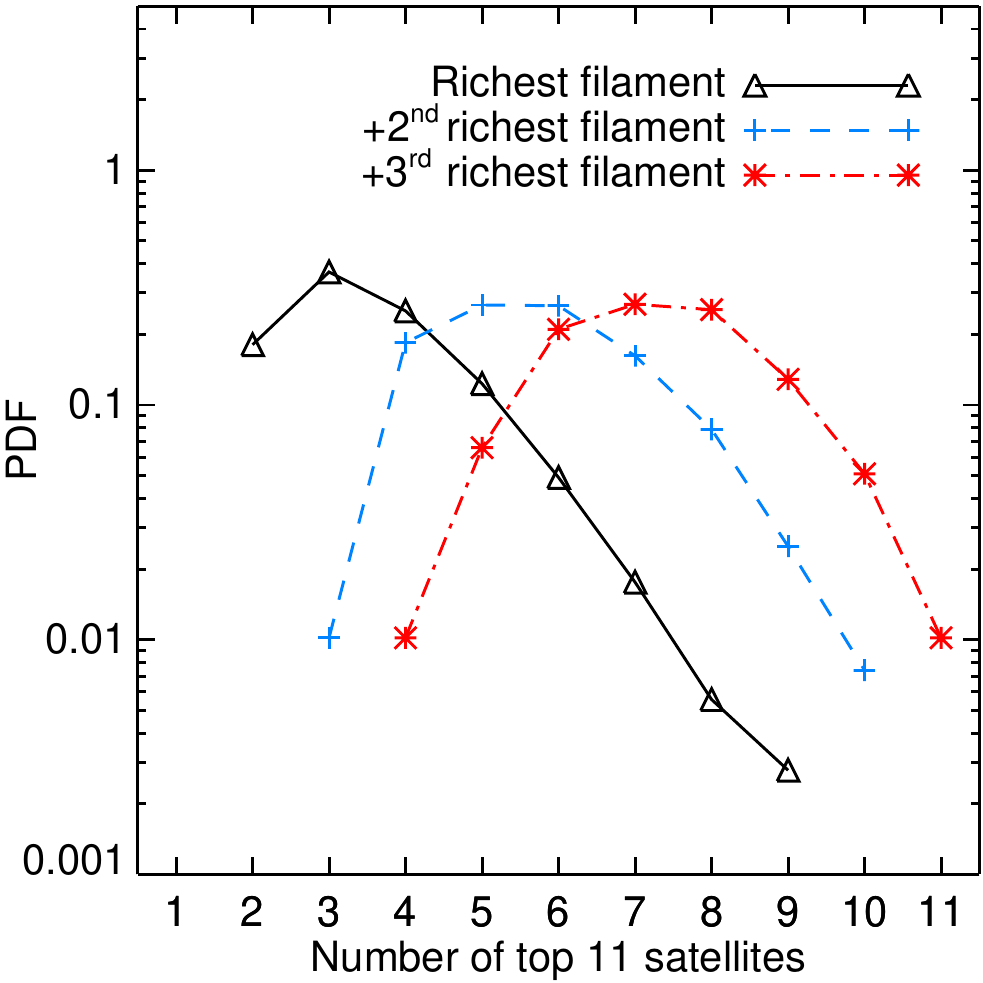}
 \caption{The PDF of the number of top 11 satellites accreted along
the richest (solid black), the two richest (dashed blue) and
three richest (dotted red) filaments in the \eagle{} simulation. The filament richness is
given by the number of satellites accreted along that filament.
 }
 \label{fig:fila_top}
\end{figure}
Filamentary accretion of satellites is an ubiquitous feature of
structure formation within $\Lambda$CDM, and, similarly to group
accretion, enhances the spatial and orbital anisotropies of the
satellite distribution. The filaments act as channels that transport
dwarf galaxies and that funnel their infall into MW-sized haloes
\citep[e.g.][]{Libeskind2005,Libeskind2014,Buck2015,Gonzalez2016}. Here,
we consider that two satellite galaxies were accreted along the same
filament if they entered their host halo along approximatively the
same direction. This definition is motivated by two
observations. First, the prominent massive filaments remain quite
stable in time, with most of the evolution of the filamentary network
involving only the thin tenuous filaments
\citep{Rieder2013,Cautun2014c}. The massive filaments are the main
mass accretion pathways into the halo and thus they are the ones along
which most satellites fall in \citep{Danovich2012}. Secondly,
filamentary accretion is more likely to lead to co-planar orbits if
two satellites enter their host halo at roughly the same points.

The more massive a satellite is, the more likely it is that it was
accreted along the spine of a filament \citep{Libeskind2014}. This
suggests a simple algorithm for identifying how many of the
present-day satellites were accreted along the same filament,
i.e. along the same direction. Starting with the most massive
satellite, we compute the angle between its entry direction and that
of the remaining top $N$ satellites. Then, all the satellites within
an opening angle of $30^{\circ}$ are assigned to the first
filament. We then go to the next most massive satellite that is yet
to be assigned to a filament and compute the angle between its entry
direction and that of the other satellites not assigned to
filaments. The second filament contains all the satellites within the
same opening angle of $30^{\circ}$. We iteratively apply this
procedure until all satellites are assigned to one filament. 
Similarly to group accretion, filamentary accretion is defined using
the satellite entry points in the host FOF halo. We 
checked that we obtain mostly the same filaments if instead we 
use the satellite entry points measured on a uniform sphere outside 
the host halo. We refer
to the number of satellites associated with each filament as the
filament richness, and we order the filaments in decreasing order of
their richness. The richest filaments are likely to correspond to the
prominent filaments feeding MW-mass haloes, with each MW-mass halo
having at least two or three such objects
\citep{Danovich2012,Cautun2013,Cautun2014c,Gonzalez2016}.

The choice of $30^\circ$ opening angle corresponds to the typical
angular size of dwarf groups at infall as seen from the centre of the
MW-mass host halo. This ensures that if the primary dwarf galaxy of
that group is assigned to a certain filament, then all the other
group members are also assigned to that filament. The $30^{\circ}$
opening angle corresponds to a solid angle of $0.27\pi$ and represents
$\sim 1/15$ of the full sky. Thus, this opening angle is small enough,
such that, if the top 11 satellites were accreted isotropically, no
two satellites would have to be part of the same filament.

\reffig{fig:fila_top} shows the satellite galaxies count brought in by
the top $i \rm{th}$ ($i=1,2,3$) richest filaments in the \eagle{}
simulation. The distribution of the richest filament shows that
for all hosts the richest filament contains two or more galaxies that
fell along it. Thus, in none of the hosts, were the satellites
accreted from 11 directions separated by more than $30^\circ$
each. The richest filament is most likely to contain three satellites
(40\% of cases), followed by four satellites in 25\% of the cases. In
20\% of the hosts five or more satellites are accreted along the richest
filament. The curve for the top two richest filaments peaks at a value
of 5$-$6 suggesting that typically half of the top 11 satellites were
accreted along just two filaments, with 80\% of hosts having accreted
at least five satellites along those two filaments. Furthermore, 70\% of
the hosts have at least seven satellites in their top three richest
filaments. If satellite accretion directions were distributed
isotropically, only 45\% of hosts would have similarly rich top three 
filaments (see Appendix \ref{appendix:filament_distribution} for details), 
which illustrates the anisotropic and filamentary nature of dwarf galaxy accretion.

\begin{figure*}
 \centering
 \begin{tabular}{cc}
 	\multicolumn{2}{c}{ \includegraphics[width=0.57\paperwidth]{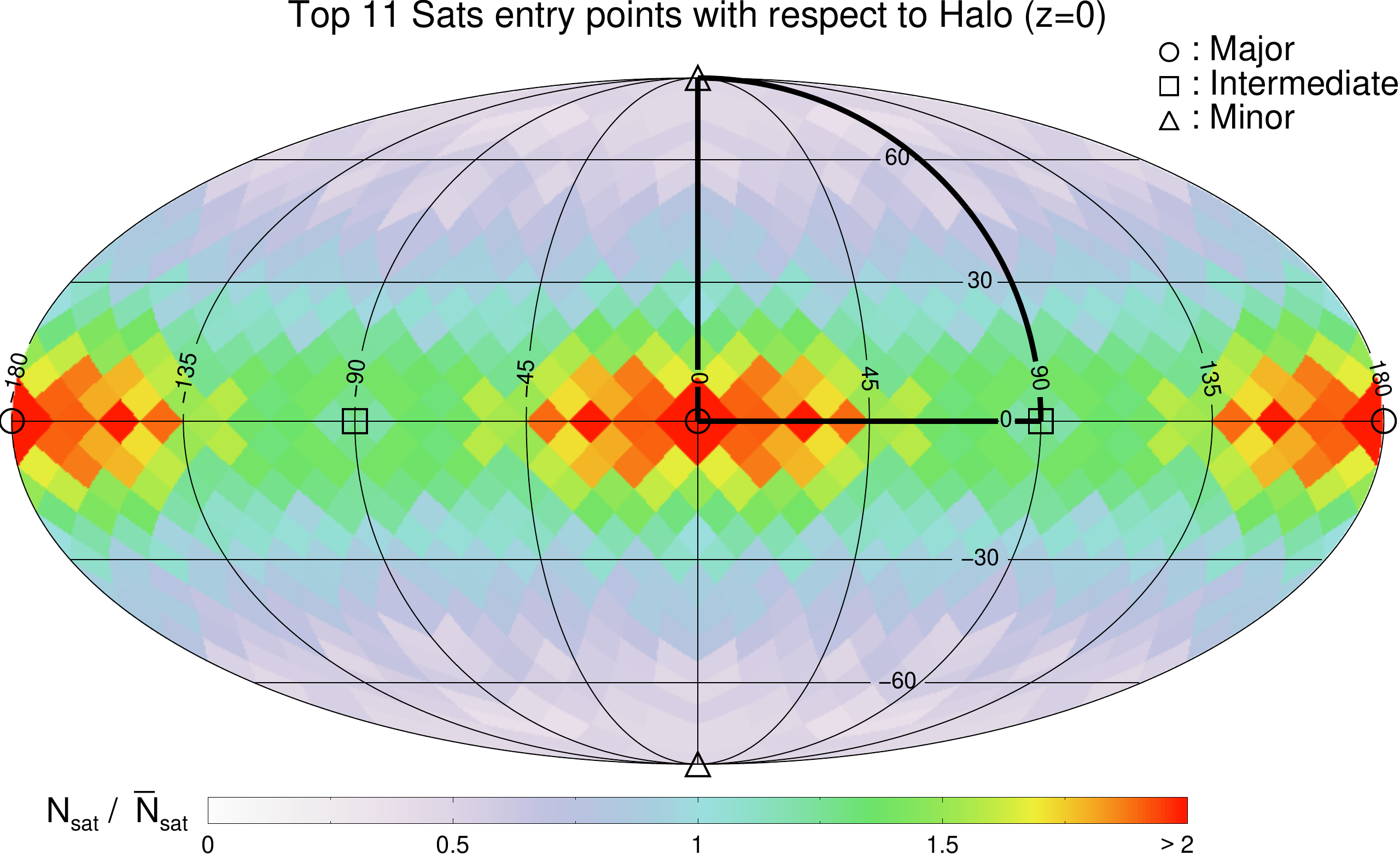} } \\[.3cm]
 	\includegraphics[width=0.4\paperwidth]{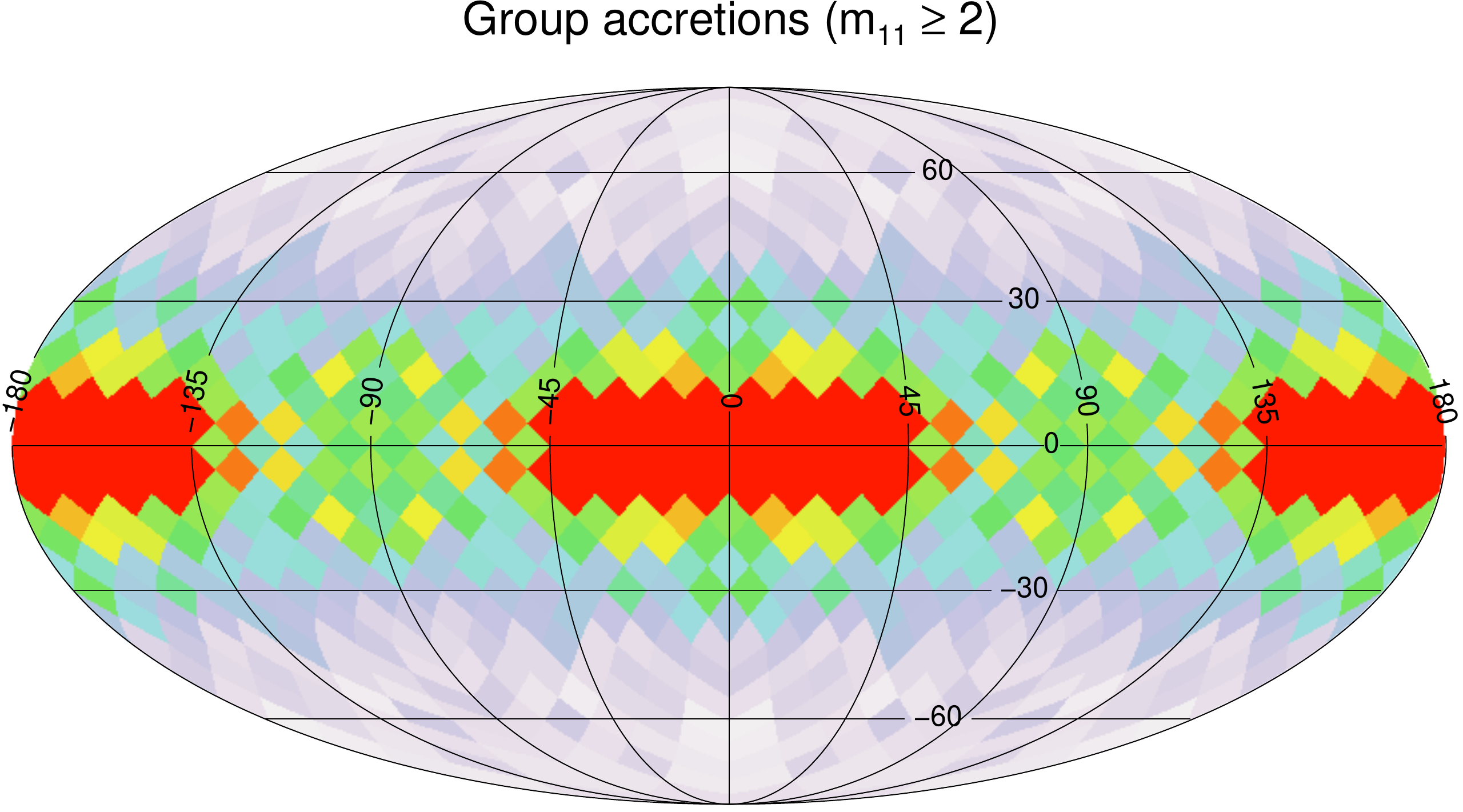} &
 	\includegraphics[width=0.4\paperwidth]{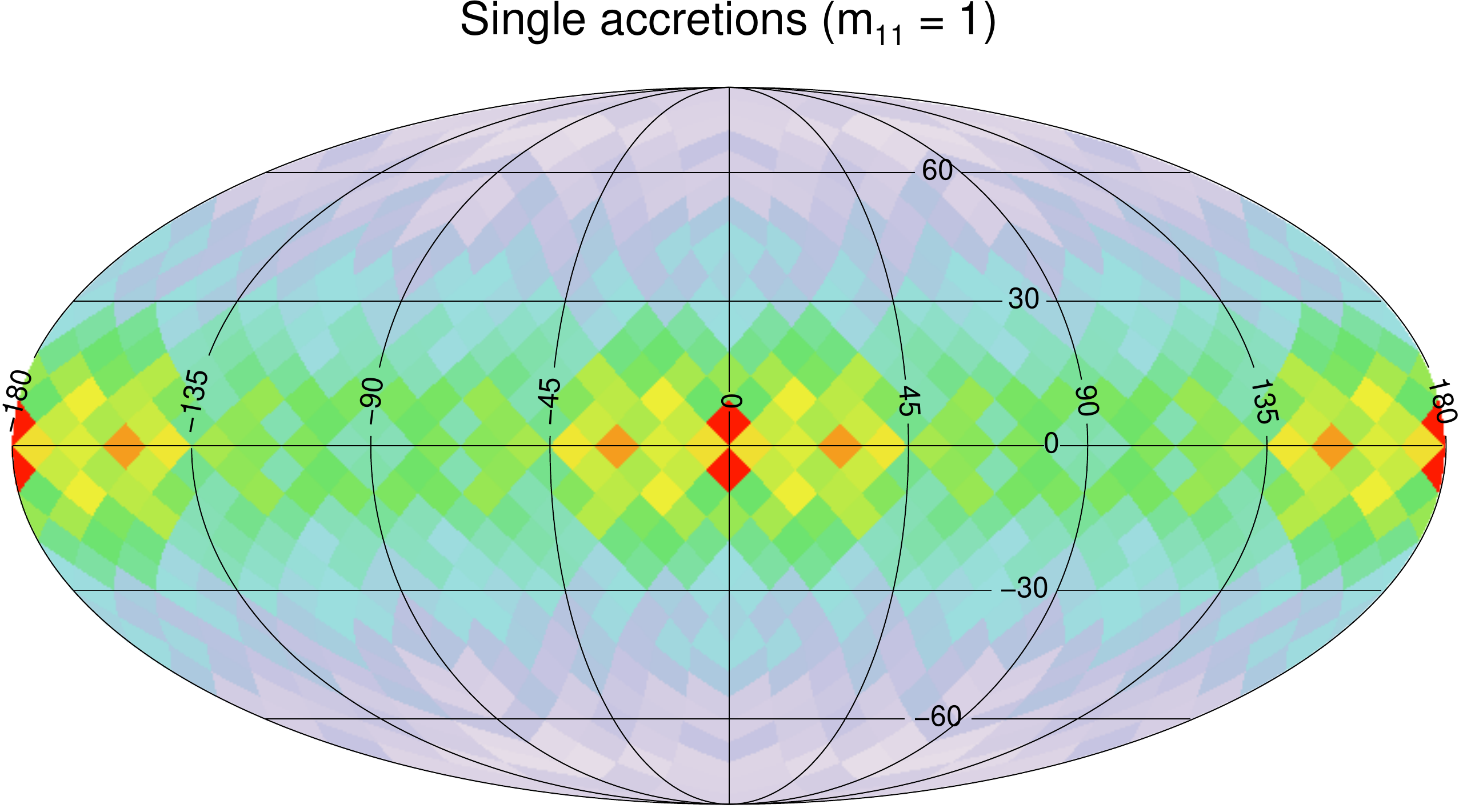} \\[.3cm]
 	\includegraphics[width=0.4\paperwidth]{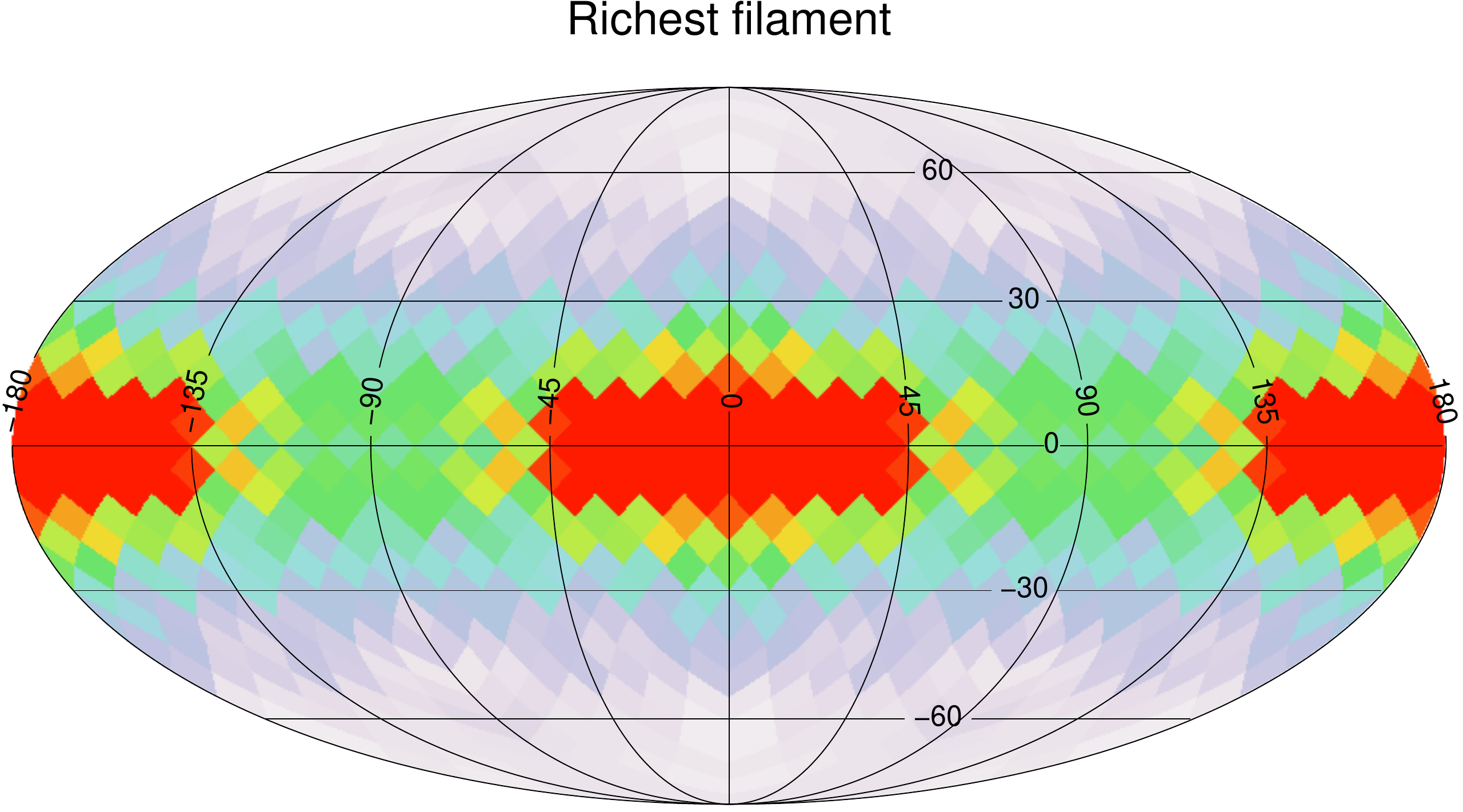} &
 	\includegraphics[width=0.4\paperwidth]{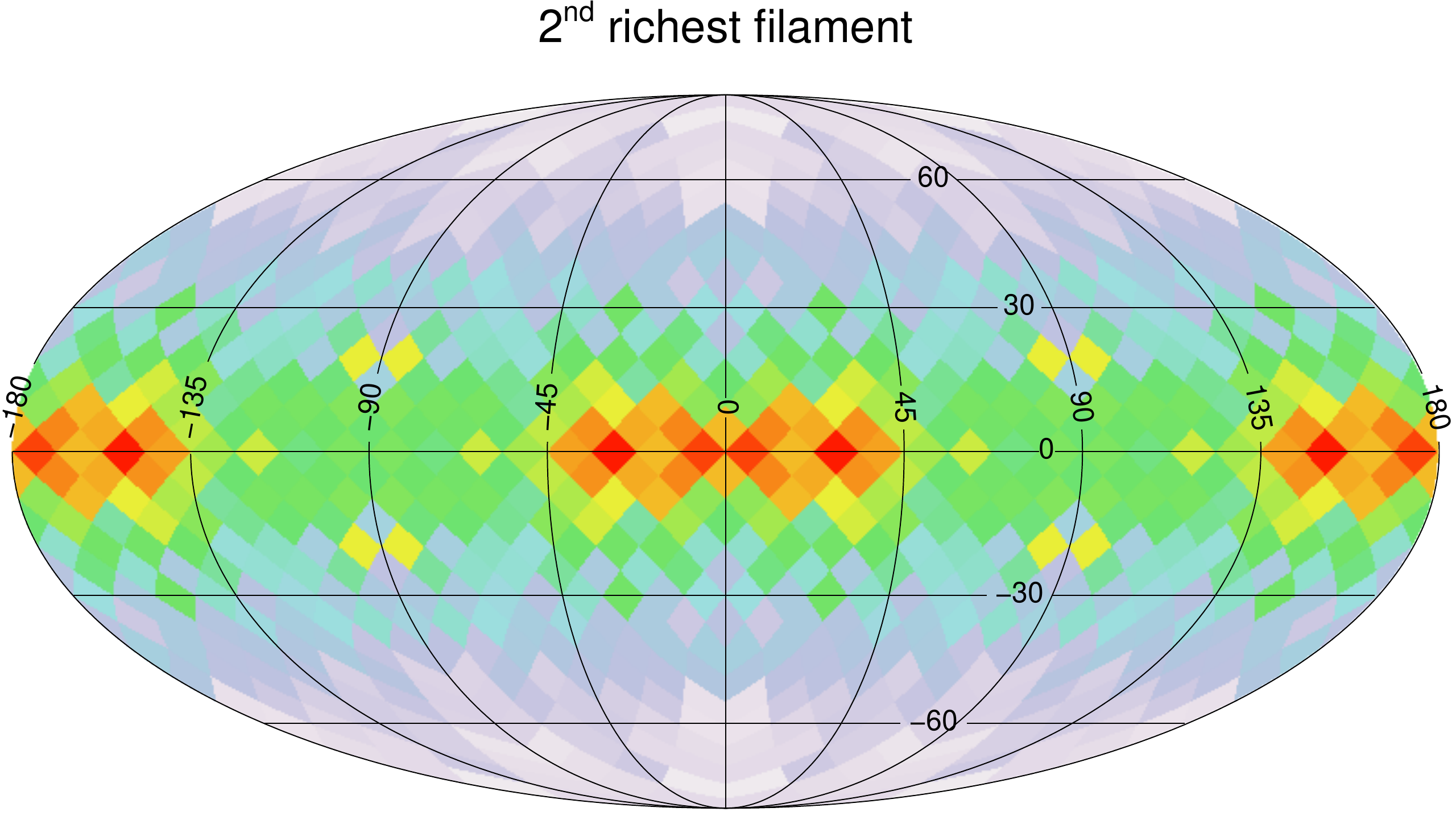} 
 \end{tabular}
 \caption{ Aitoff sky projections of the entry points of the top 11
satellites in the \eagle{} simulation. To stack all the hosts, we
expressed the entry point in a coordinate system given by the
preferential axes of the $z=0$ DM halo shape, with the main,
intermediary, and minor axes corresponding to the points $(0,0)$,
$(90,0)$ and $(0,90)$ degrees, respectively 
(see the legend of the top panel). The colour scale is the
same for all panels and shows the number of satellites in \healpix{}
pixels normalized by the mean expectation for isotropic accretion. The
halo shape determines an orientation, not a direction, which means
that all the relevant information is contained in one octant of the
sphere (the thick solid line in the top panel shows one such octant);
this explains the symmetries of the projection. The top panel
shows the entry points of all satellites, the middle one shows the
entry points of group (left) and single (right) accretion events, and
the bottom one shows the positions of the richest (left) and second
richest (right) filaments.}
 \label{fig:sky}
\end{figure*}

\subsection{Anisotropy of accretion}
\label{Sec:ani}
\begin{figure}
 \plotone{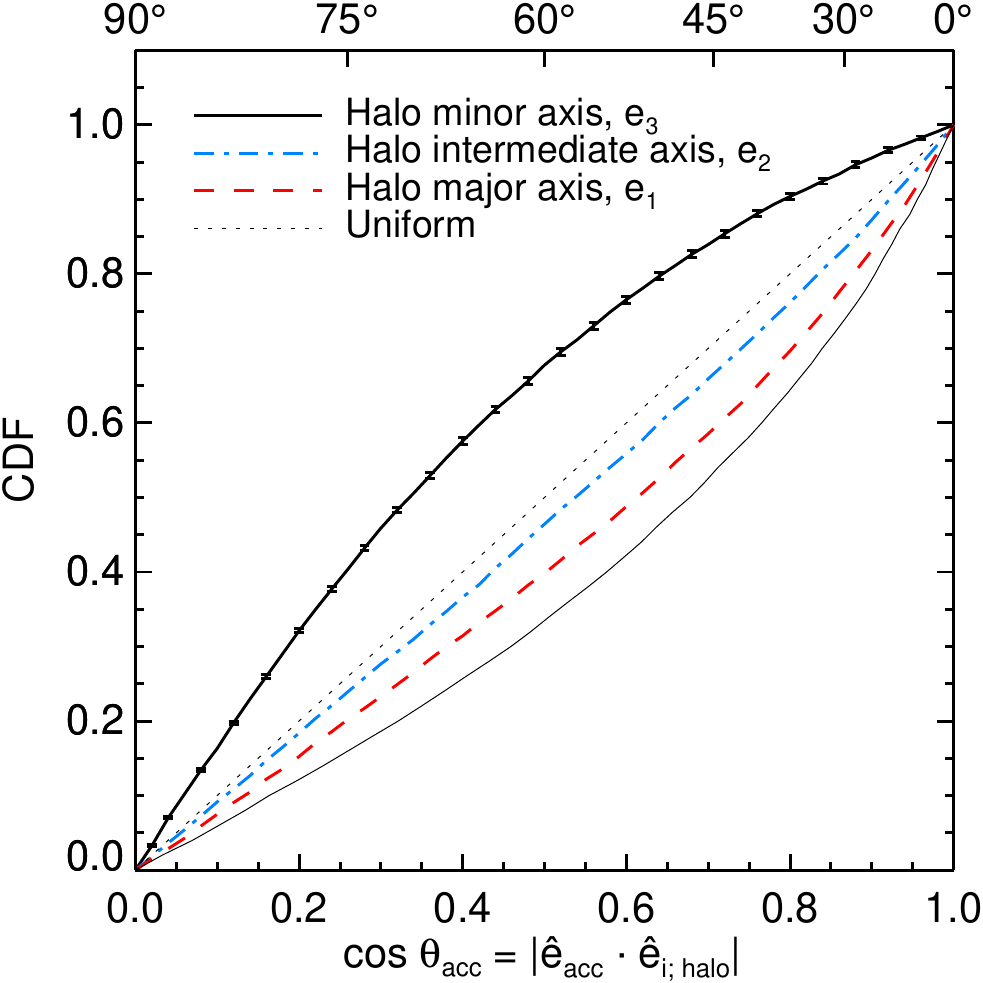}
 \caption{ The cumulative distribution function (CDF) of the accretion
misalignment angle, $\theta_{\rm acc}$, between the entry points of
the top 11 satellites and the major (dashed line), intermediate
(dash-dotted line) and minor axes (thick solid line) of the shape of
their $z=0$ host haloes in the \eagle{} simulation. The error bars around the thick solid line
indicate the $1\sigma$ uncertainty range, which has the same size for
all the curves. The thin dotted line corresponds to a uniform
distribution. The thin solid line is the mirror image of the thick
solid line with respect to the uniform line; it shows that the largest
alignment is with the halo minor axis.
The error bars shown here and in the subsequent figures show the 68
percentile bootstrap uncertainties calculated using 200 samples.
 }
 \label{fig:theta_halo}
\end{figure}

\begin{figure}
 \plotone{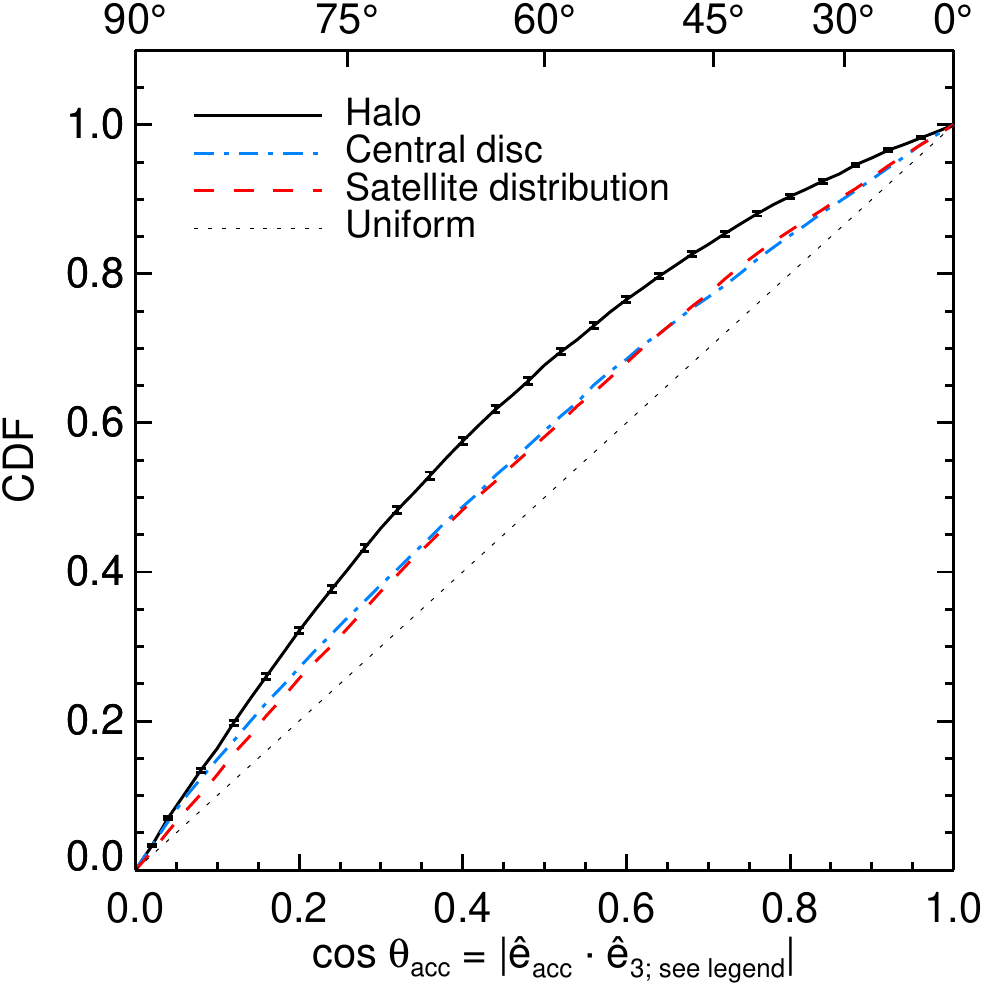}
 \caption{ The CDF of the accretion misalignment angle, $\theta_{\rm
acc}$, between the entry points of the top 11 satellites and the shape
minor axis, $e_3$, of their $z=0$ host haloes (solid line), central
galaxies (dash-dotted line), and top 11 satellite distributions
(dashed line) in the \eagle{} simulation.
 }
 \label{fig:theta_plane}
\end{figure}

\begin{figure}
 \plotone{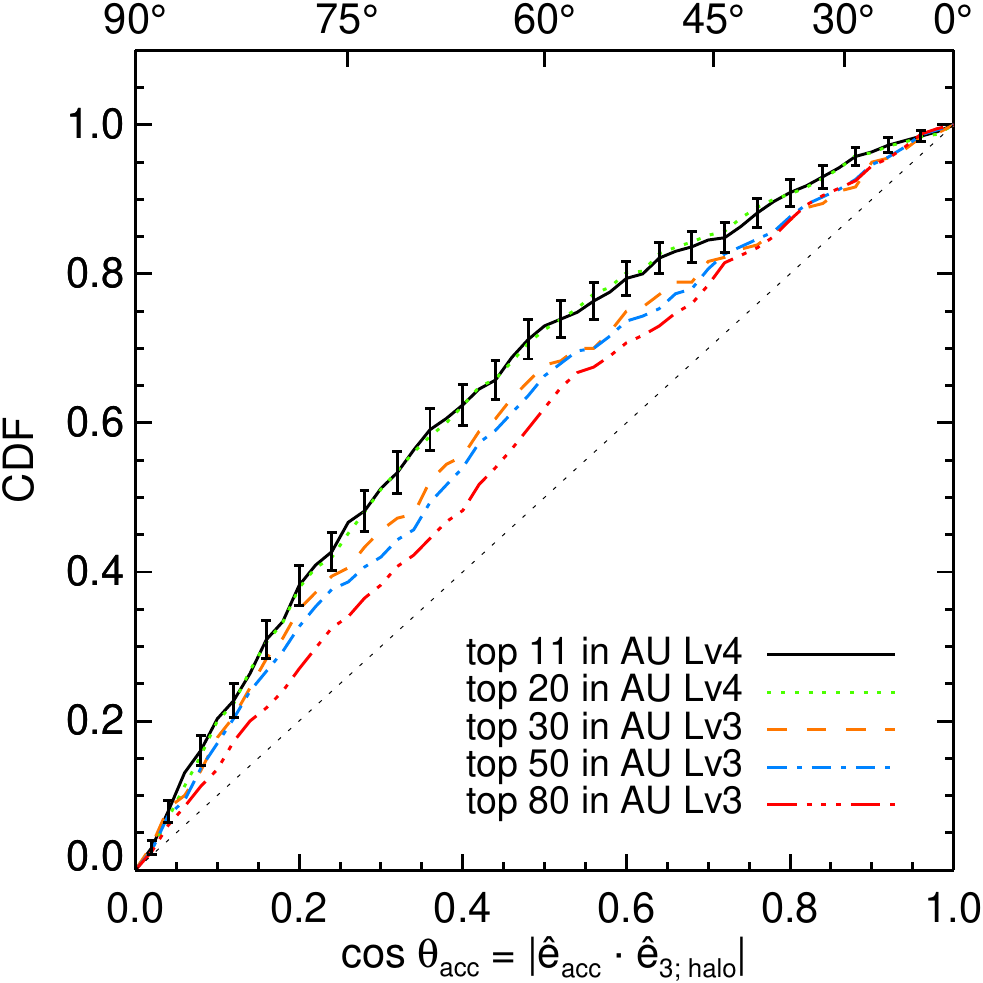}
 \caption{ The CDF of the accretion misalignment angle, $\theta_{\rm
acc}$, between the entry points of the top $N$ satellites and the
shape minor axis of their $z=0$ host haloes. The various curves show
the dependence of the CDF on $N$ in the \auriga{} medium- (level-4) and
high-resolution (level-3) simulations. The \eagle{} top 11 results
agree well with the \auriga{} level-4 top 11 ones and, for
readability, are not shown in the figure.}
 \label{fig:theta_plane_vary_top_N}
\end{figure}

To study the anisotropy of satellite accretion in the \eagle{}
simulations, we examine the entry points of the top 11 satellites. As
described in \refsec{sec:multi}, starting at high redshift, we trace
the top 11 satellites forward up to the first simulation output when
they become part of the same FOF group as the progenitor of their
present day MW-mass host; this determines the accretion time. Then,
using the snapshot just before accretion, we define the entry point of
each satellite as the position of the satellite progenitor with
respect to the central galaxy progenitor.

We calculate the accretion anisotropy with respect to the $z=0$ shapes
of the DM host halo, central galaxy and the top 11 satellite
distribution. The shape is determined from the mass tensor,
\begin{equation}
 I_{ij} \equiv \sum_{k=1}^{N} m_{k} x_{k,i} x_{k,j} \;,
 \label{eq:inertia_tensor}
\end{equation}
where $N$ is the number of tracers, which are DM particles for the
halo, stars for the central disc and the top 11 satellites for the
satellite distribution. For the halo we limit our calculation to all
particles included within $\rm R_{200}$, while for the central galaxy
we use all the stars within $10\kpc$ from the galaxy centre. The
quantity $x_{k,i}$ denotes the $i$th component ($i=1,2,3$) of the
position vector of tracer $k$ with respect to the halo centre, and
$m_k$ denotes that tracer's mass. For the shape of the satellite
population, we weigh equally all satellites by assigning them the same
mass. The shape and the orientation are determined by the eigenvalues,
$\lambda_i$ ($\lambda_{1}\geqslant\lambda_{2}\geqslant\lambda_{3}$),
and the eigenvectors, $\uvec{e}_i$, of the mass tensor. The major,
intermediate and minor axes of the corresponding ellipsoid are given
by $a=\sqrt{\lambda_1}$, $b=\sqrt{\lambda_2}$, and
$c=\sqrt{\lambda_3}$, respectively.

In \reffig{fig:sky}, we show the entry points of the top 11 satellites
of \eagle{} MW-sized hosts. To stack all the hosts, we expressed the
entry directions into a common coordinate system, which we chose as
the eigenvectors of the shape of the DM halo. As we will discuss
shortly, satellite accretion shows the strongest alignment with the
halo shape, which represents the motivation for our choice of
coordinate system. The eigenvectors determine only an orientation and
do not have a direction assigned to them, so, when expressed in the
halo shape eigenframe, we mirror each entry point eight times: two times
for each Cartesian coordinate. This means that independent information
is contained in only one octant of the sky plot; however, for clarity,
we choose to present the full sky distribution. By stacking the top 11
satellites of all the $1080$ MW-sized hosts, we have a sample of
$11880$ entry points which allows for a statistically robust
characterization of dwarf galaxy accretion anisotropies. The 3D entry
direction of each satellite is expressed in spherical angular
coordinates and pixlized using the \healpix{} public
code. \footnote{http://healpix.sourceforge.net}

The top row in \reffig{fig:sky} shows that satellite accretion is
highly anisotropic, with dwarf galaxies being preferentially accreted
within $20^\circ$ from the equatorial plane, which is determined by
the major and intermediate axes of the DM halo, and, within this
plane, satellite accretion shows a pronounced excess along the halo
major axis. This not only confirms the anisotropic accretion results
found by previous studies
\citep{Libeskind2011,Libeskind2014,Kang2015,Wang2018}, but extends those
results to hydrodynamic simulations \citep[see][]{Garaldi2018} and 
to a large sample of MW-sized
haloes, which allows us to robustly quantify anisotropic
accretion. The marked alignment between halo shape and satellite
infall is due to both mass and satellites being preferentially
accreted along a few massive filaments; this filamentary infall is the
one giving rise to the alignment of both DM halo shape
\citep{Zhang2009} and satellite galaxies \citep{Tempel2015} with the
cosmic web filaments.

The middle row in \reffig{fig:sky} illustrates how the anisotropy of
accretion for the top 11 satellites varies between group and singly
accreted dwarfs. Compared to single accretion events, galaxies that
arrive with one or more companions (20\% of the population) are more
strongly clustered along the equatorial plane of the projection and
especially along the halo major axis. This is to be expected, since on
average groups of dwarfs reside in more massive subhaloes than single
dwarfs at accretion, and more massive subhaloes are more likely to be
accreted along filaments than less massive ones
\citep{Libeskind2014}. This strong correlation between group and
filamentary accretion is emphasized by the bottom row of
\reffig{fig:sky}, which shows the entry points of the dwarfs that were
accreted along the richest and second richest filaments. The
satellites that fell in along the richest filament (42\% of the
population) are strongly clustered along the halo major axis, very
similar to the clustering of group accretions. The second richest
filament is preferentially located within the equatorial plane and
roughly randomly oriented within this plane. 

To better quantify the anisotropy of satellite accretion, we define
the accretion misalignment angle, $\theta_{\rm acc}$, between the
satellite entry point and the present-day DM halo, central disc or
satellite distribution. The misalignment angle is given by
\begin{equation}
 \cos\theta_{\rm acc} = |\uvec{e}_{\rm acc}\cdot\uvec{e}_{i;~\rm{X}}| \;,
 \label{eq:misalignment_angle}
\end{equation}
where $\uvec{e}_{\rm acc}$ is the unit vector pointing along the
satellite entry point, and $\uvec{e}_{i;~\rm{X}}$ ($i=1,2,3$) are the
principal axes of the $z=0$ shape of the: DM halo ($\rm{X=halo}$),
central disc ($\rm{X=disc}$) or satellite distribution
($\rm{X=sats}$).

We start by studying the misalignment angle between satellite
accretion and present-day DM halo shape, which is shown in
\reffig{fig:theta_halo}. Satellite accretion tends to be well aligned
with the major axis (median angle $52^\circ$), to be less aligned with
the intermediate axis (median angle $57^\circ$), and to be
preferentially perpendicular to the minor axis (median angle
$70^\circ$). To compare the three signals, we mirror the minor axis
alignment with respect to the expectation for an isotropic
distribution, which is the diagonal line. We find that the accretion
direction of satellites shows the largest alignment (actually a
misalignment) with the halo minor axis, which suggests that the most
important trend is for satellites to fall in perpendicular to the halo
minor axis. We studied the misalignment angle between accretion
direction and the $z=0$ central disc and satellite distribution, and,
in both cases, the strongest alignment is with the minor axis, which
is why in the following figures we choose to show only the alignment
angle with respect to the minor axis.

\reffig{fig:theta_plane} compares the alignment of the satellite
infall directions with the $z=0$ shape minor axis of the halo, central
galaxy and the top 11 satellites. Of the three, the halo minor axis
shows the largest alignment (median angle $71^\circ$), while the
central galaxies and the satellite systems show lesser degree of
alignment (both have median angles of $66^\circ$). While central
galaxies are well aligned to the innermost ${\sim}20\kpc$ of their DM
haloes \citep{Shao2016,Gomez2017a}, the galaxies show on average a
$33^\circ$ misalignment angle with the full DM distribution within
$R_{200}$ \citep{Shao2016}. The misalignment could be due to time
variations in position of the filaments along which most matter is
accreted into the halo \citep{Vera-Ciro2011,Rieder2013} and due to
massive substructures that can torque the inner disc
\citep{Gomez2017a,Gomez2017b}. Similarly, the satellite distribution
also shows a $33^\circ$ misalignment angle with the full DM halo
\citep{Shao2016}, which could be due to the satellites representing a
stochastic sampling of the DM distribution \citep{Hoffmann2014}.
\citet{Shao2016} pointed out that the present-day central galaxy --
satellite system alignment is a consequence of the tendency of both
components to align with the DM halo. This could also be the case for
the alignment of the infall directions. For example, the alignment of
the central galaxy with the halo, which in turn is aligned with the
satellite infall direction, would result in a weak alignment between
the central galaxy and the satellite infall direction, which is what
we measure.

In \reffig{fig:theta_plane_vary_top_N} we study how the anisotropy of
accretion varies as a function of satellite brightness using the
\auriga{} medium- and high-resolution simulations. The degree of
anisotropic accretion decreases from bright to faint satellites, for
example the median misalignment angle of the top 11 satellites is
$73^\circ$ while for the top 80 satellites is $66^\circ$. This agrees
and extends the results of \citet{Libeskind2014}, who have shown that
the most massive dark matter subhaloes are the ones that were accreted
most anisotropically.

\begin{figure}
 \plotone{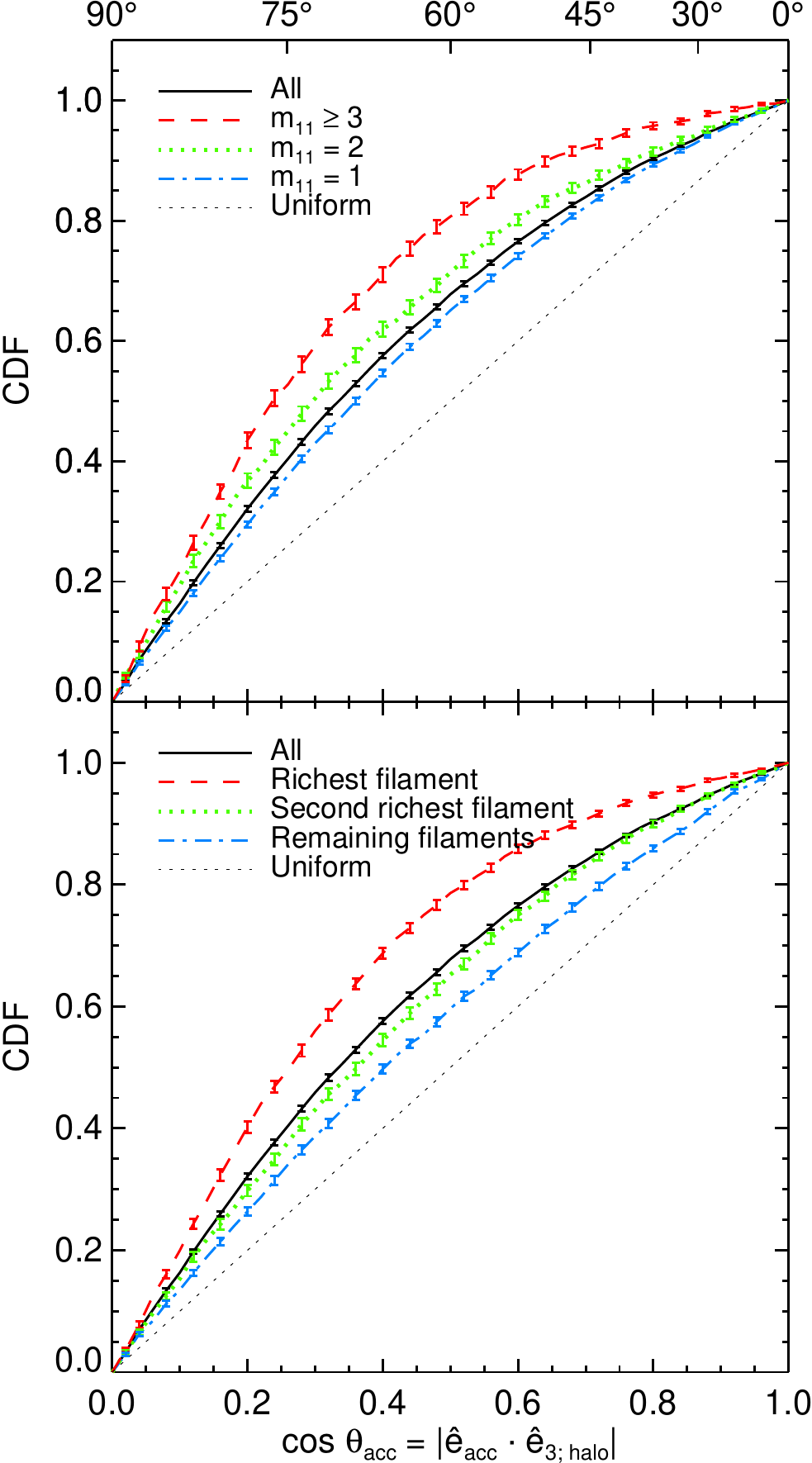}
 \caption{ The CDF of the accretion misalignment angle, $\theta_{\rm
acc}$, between the entry points of the top 11 satellites and the shape
minor axis of their $z=0$ host haloes in the \eagle{} simulation.
The black line shows the full
sample alignment while coloured lines correspond to subsamples. In
the upper panel, satellites are split according their multiplicity of
accretion, $\rm m_{11}$, while in the bottom panel satellites are
split according to the richness of the filament they fell along. The
error bars indicate the $1\sigma$ uncertainty for the various
subsamples.
 }
 \label{fig:theta_mul}
\end{figure}

\begin{figure}
 \plotone{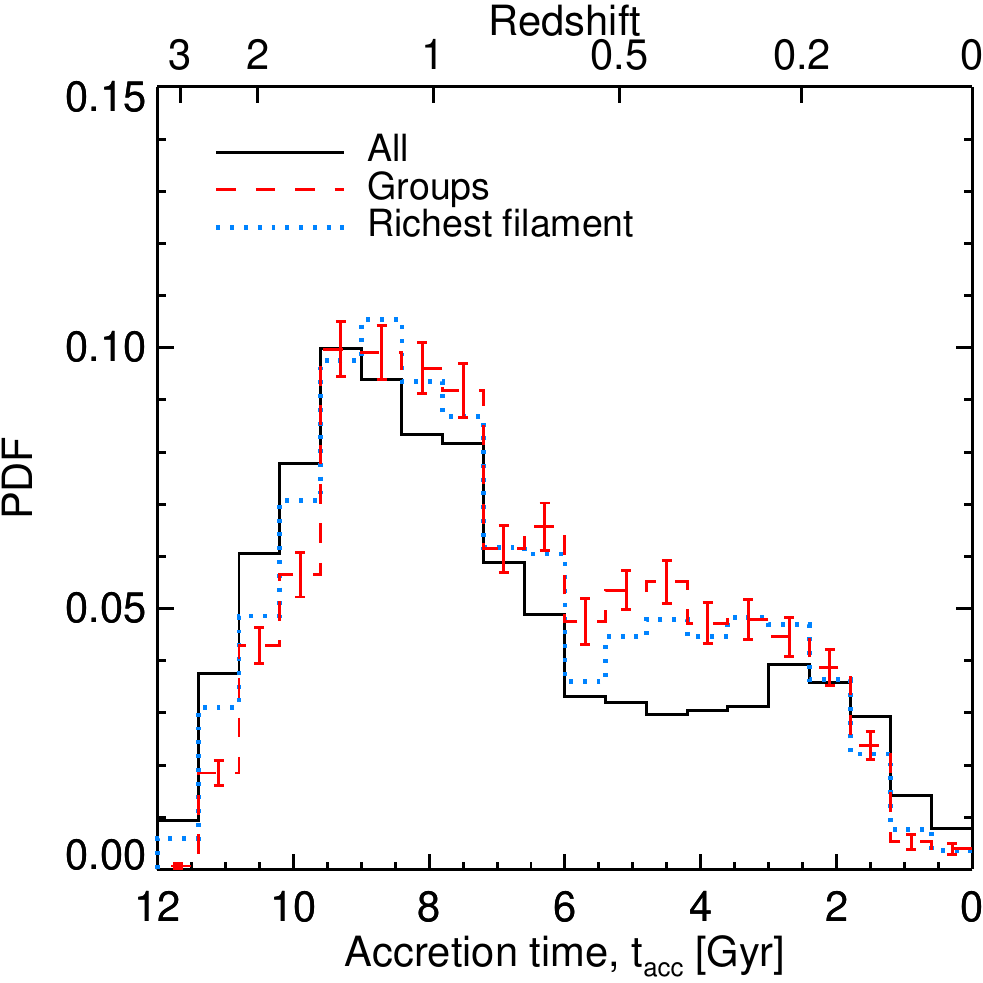}
 \caption{ The distribution of top 11 satellites' accretion lookback
time, $t_{\rm acc}$, in \eagle{} for the full sample (solid line), and
for two subsamples: satellites that were accreted in groups, i.e. $\rm
m_{11} \geq 2$ (dashed line), and those that were accreted along the
richest filament (dotted line). The error bars indicate the $1\sigma$
uncertainty range, which, for clarity, we show only for the subsample
with the smallest size: satellites accreted in groups.
 }
 \label{fig:infall}
\end{figure}

In \reffig{fig:sky}, we found that satellites accreted in groups and
those along the richest filaments show a larger degree of anisotropic
infall than the whole population. We further quantify this effect in
\reffig{fig:theta_mul}, where we present the misalignment angle
between infall direction and the DM halo minor axis for a subsample of
satellites selected according to their group multiplicity (top panel)
or filament richness (bottom panel). Rich groups, that is with a
multiplicity of at least 3, show a much larger misalignment than the
full satellite population. A similar trend is observed for pair
accretion too, though in this case the difference with the full sample
is smaller. Singly accreted satellites show roughly the same
alignment, albeit slightly weaker, than the full sample. The entry
points of satellites that fell in along the richest filament are more
anisotropic than the whole population, while dwarfs associated with the
second richest and lower richness filaments have a similar anisotropy,
which is weaker than that of the full sample.

Satellites that fell in as part of a group or along the richest
filament have on average a later accretion time than the full
sample. This is illustrated in \reffig{fig:infall}, which shows the
distribution of accretion times, with $t=0$ corresponding to present
day. The accretion time distribution is wide, with the full sample of
top 11 satellites having a typical accretion time, $t=8.5$
Gyr. Compared to the whole sample, satellites accreted in groups and
along the richest filament have systematically late accretion time,
with few objects accreted before $t=11$ Gyr and a considerable excess
of objects with $t\leq6$ Gyr. Multiply accreted satellites are more
likely to have fallen in inside massive haloes, which, since those
haloes needed time to grow, could explain why they were accreted
later. Galaxies falling in along filaments are more likely to be on
radial orbits and thus more likely to be disrupted
\citep{Gonzalez2016}, and thus the surviving satellites are more
likely to be recently accreted ones that were not inside their $z=0$
MW hosts for enough time to experience significant tidal
stripping. While not shown, we also find a slight tendency for late
accreted satellites to have infall directions that are more aligned
with the present-day halo than early accreted objects. But this trend
is not strong enough to explain the results of \reffig{fig:theta_mul},
that is the larger accretion anisotropy of satellites associated with
groups and to the richest filament.

\subsection{Implications for the MW disc of satellites}
\label{sec:cons}

\begin{figure}
 \plotone{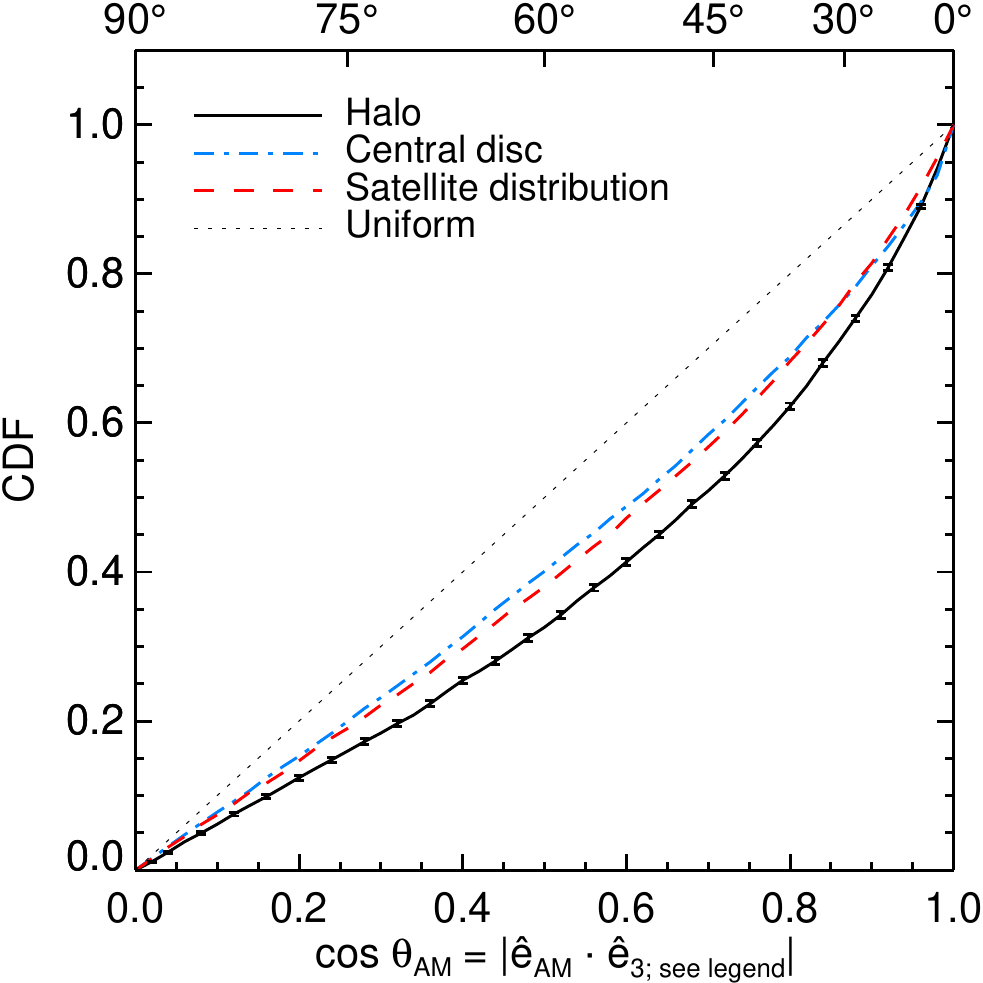}
 \caption{ The CDF of the orbital pole misalignment angle,
$\cos(\theta_{\rm AM})$, between the $z=0$ orbital angular momentum
vector of the top 11 satellites and the shape minor axis, $e_3$, of
their $z=0$ host haloes (solid line), central galaxies (dash-dotted
line), and top 11 satellite distributions (dashed line) in the \eagle{} simulation.
 }
 \label{fig:theta_am}
\end{figure}

\begin{figure}
 \plotone{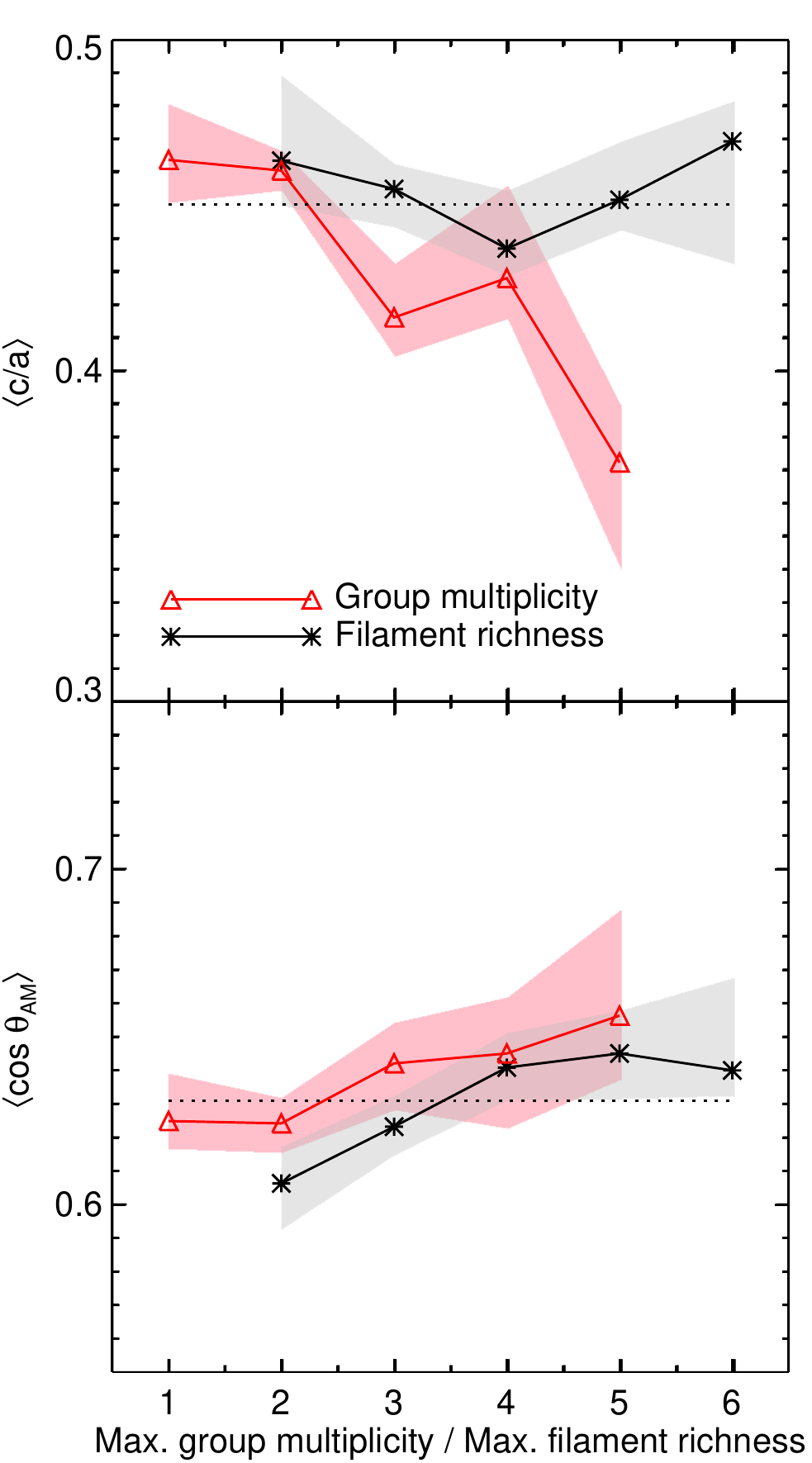}
 \caption{ Upper panel: the median flattening, $c/a$, of the top 11
satellite distributions as a function of the maximum group
multiplicity (red triangles) and the maximum filament richness (black
squares) of the host halo. Bottom panel: the median orbital pole
angle, $\cos(\theta_{\rm AM})$, between the satellite orbital angular
momentum and the minor axis of the satellite distribution as a
function of maximum group multiplicity and maximum filament
richness. Solid lines and shaded regions indicate the median values of
the distribution and the $1\sigma$ error in determining the median,
respectively. The horizontal dotted line indicates the median value
for the whole sample.
 }
 \label{fig:am-mul}
\end{figure}

Group and filamentary accretion has been suggested as an explanation
for the MW disc of satellite galaxies
\citep[e.g.][]{Libeskind2005,Libeskind2011,Li2008,Lovell2011} which
consists of two main features: a very flattened spatial distribution
and a large clustering of the orbital poles. Our large sample of
\eagle{} MW-mass haloes offers the perfect opportunity to check this
conjuncture, which is the aim of this section.

\citet{Shao2016} studied the flattening of the top 11 satellite
distribution for the same sample of \eagle{} MW-mass haloes used
here. The flattening, $c/a$, is defined as the ratio of the minor to
major axes of the satellite distribution. The top 11 satellites in
\eagle{} have a large spread in $c/a$ values, with a median value
$c/a \approx 0.45$ (see Fig. 2 in \citealt{Shao2016}). In contrast,
the MW classical dwarfs have $c/a=0.183\pm0.008$, which is in the tail
of the \eagle{} distribution, with only ${\sim}1\%$ of simulated
systems having a flattening at least as extreme as the one observed
for the MW. This is in general agreement, although slightly lower than
previous studies \citep{Wang2013,Pawlowski2014b,Pawlowski2014c}. 
The difference could be due to tidal stripping of satellites 
by the baryonic disc of the central galaxy, which leads to less 
concentrated radial distributions of satellites 
\citep{Ahmed2017,Sawala2017} and thus less flattened satellite
distributions.

In \reffig{fig:theta_am}, we present the alignment between the $z=0$
satellite orbital poles and the minor axis of the DM halo, central
galaxy and top 11 satellite distribution. Although the orbital poles
align with the central disc and the satellite system, the largest
alignment is with the halo minor axis (median angle of $46^\circ$),
which indicates that dwarf satellites preferentially orbit in a plane
perpendicular to the halo minor axis. This result qualitatively agrees
with previous literature, which studied the orbital pole -- halo minor
axis alignment using subhaloes \citep{Lovell2011} or using
semi-analytical galaxy formation models \citep{Cautun2015b}; a more
quantitative comparison is difficult since the median alignment angle
depends on the mass limit used to select the satellite sample, with
faint satellites having more misaligned orbits (see Fig. 4 in
\citealt{Cautun2015b}).

The majority of the classical MW satellites have orbital poles
pointing along the normal to the Galactic plane of satellites
\citep{Pawlowski2012b}, with the median orbital pole -- satellite
system misalignment angle being $\theta_{\rm AM}=36^\circ \pm
6^\circ$ \citep[we obtained this value using the MW satellite
positions and velocities given in][]{Cautun2015}. In \eagle{}, the
same misalignment angle has a median value $\theta_{\rm
AM}=51^\circ$; this is significantly less aligned than the MW value
and illustrates the enhanced orbital pole clustering seen for the
classical MW dwarfs.

The upper panel of \reffig{fig:am-mul} shows the correlation between
the richest accreted satellite group and the median $c/a$ ratio, which
we calculated by splitting the MW-mass halo sample according to the
maximum group multiplicity of each host. We find 
that groups with a multiplicity of 3 or 4 already lead to slightly 
thinner planes, but the effect becomes especially important for groups with a
multiplicity of 5 or higher. Nevertheless, accretion of such rich groups is rare, with only
4\% of systems having accreted a group with multiplicity of 5 or higher.
The same panel also shows the relation
between the median $c/a$ value and the maximum filament richness of a
host halo to find that the two are uncorrelated. It suggests that, if
filamentary accretion is the explanation for flat satellite distributions, it is
not simply due to many satellites falling in along the same filament,
and, probably, the decisive factor is the overall configuration of
filaments (see Appendix \ref{sect:appendix_two}), whose study we leave for
future work. Alternatively, it has been suggested that rotating planes of satellite 
galaxies could be due to other processes, such as tidal dwarf galaxies \citep[e.g. see][]{Fouquet2012,Hammer2013}.

The bottom panel of \reffig{fig:am-mul} shows the dependence of the
median orbital pole -- satellite system misalignment angle on both
maximum group multiplicity and maximum filament richness of the
MW-mass hosts. We find that within the error bars, the results are
consistent with no dependence of the misalignment angle on either
group multiplicity or filament richness. The lack of dependence on group
multiplicity given that rich groups lead to a flatter satellite distribution
is especially puzzling (see the top panel of
\reffig{fig:am-mul}). Rich groups typically fall in inside massive
haloes, so those members galaxies can have significant velocities with
respect to each other and that, in turn, can lead to different orbital
planes inside their present-day MW-mass hosts. The satellites accreted
along the same filament, while having roughly the same entry point in
the host halo, can have different orbital momenta due to either the
filament's velocity dispersion or due to being accreted at different
times.

\section{Conclusions}
\label{sect:conclusion}
We have used two hydrodynamical cosmological simulations, \eagle{} and
\auriga{}, to study the accretion of dwarf galaxies into galactic mass
haloes. The two simulations self-consistently incorporate the main
physical processes that affect galaxy evolution and give rise to dwarf
satellite stellar mass functions that are in agreement with both MW
and M31 observations (see \reffig{fig:mf-all}). This work studied
MW-mass haloes (median mass ${\sim}1\times10^{12}\Msun$) and their
satellite population within a distance of $300\kpc$ from the central
galaxy. When applied to \eagle{}, the selection criteria resulted in
1080 MW-mass haloes that have at least 11 luminous satellites; this
constitutes our sample of MW classical satellite analogues and its
large size is ideal for a statistical study. The zoom-in \auriga{}
simulations, while having only 30 MW-sized haloes, are perfect for
studying satellites with stellar mass as low as $5\times10^4$ and
$5\times10^3\Msun$ for respectively the medium- and high-resolution
runs.

We investigated three aspects of dwarf galaxy accretion into MW-sized
haloes: the accretion of galaxy groups, the infall along the cosmic
web filaments, and the anisotropic nature of satellite
accretion. Group multiplicity was defined as the number of companion
galaxies that fell in as part of the same FOF group and that at $z=0$
are in the top $N$ largest stellar mass satellites, for varying values
of $N$. Motivated by filamentary accretion leading to similar entry
points into the host halo, filament richness was defined as the number
of top $N$ dwarfs that fell into the MW-mass host within a $30^\circ$
opening angle. The anisotropic accretion of satellites was
characterized in terms of the alignment between the satellite entry
points and the preferential axes of the $z=0$ shape of the DM halo,
central galaxy disc and the top 11 satellite distribution.

\noindent ~~~~ Our main conclusions are as follows.
\\[-.55cm]
\begin{enumerate}
 \item For the present-day top 11 satellites, 75\% of them were
accreted by themselves, 14\% in pairs, 6\% as triplets and the
rest as part of higher multiplicity groups (see \reffig{fig:frac}).
 \\[-.3cm] 
 \item Group accretion becomes more common when considering fainter
satellite samples. For example, for the present-day top 50 satellites,
60\% were accreted singly, 12\% in pairs, and 28\% in triplets
or richer groups (see \reffig{fig:frac}).
 \\[-.3cm]
 \item The multiplicity of infall groups depends on the stellar mass
of the primary (i.e. most massive) group member. LMC-sized groups,
where the primary galaxy has a stellar mass in the range
$10^9-10^{10}\Msun$, bring on average 3, 7 and 15 members with stellar
mass larger than $10^6$, $10^5$ and $10^4\Msun$ respectively. In
contrast, Fornax-sized groups (primary stellar mass in the range
$10^7-10^8\Msun$) have on average only two members more massive than
$10^4\Msun$ stellar masses (see \reffig{fig:mf-acc}). The
group-to-group variation in the stellar mass function of dwarf galaxy
groups is large, with LMC-sized groups having anywhere between $2$ and
$12$ (16 and 84 percentiles) members more massive than $10^5\Msun$.
 \item Of the $z=0$ top 11 satellites, 50\% of them are accreted
along the two most richest filaments and 70\% along the three most
richest filaments (see \reffig{fig:fila_top}).
 \\[-.3cm]
  \item Dwarf galaxy accretion is highly anisotropic, takes place
preferentially in the plane determined by the major and intermediate
axes of the DM host halo shape, and, within this plane, is clustered
along the shape major axis (see Figs \ref{fig:sky} and
\ref{fig:theta_halo}). The satellite entry points are preferentially
aligned with the central disc and the top 11 satellite system, but to
a lesser extent than the alignment with the DM halo (see
\reffig{fig:theta_plane}).
 \\[-.3cm]
 \item The degree of anisotropic accretion is largest for the most
massive satellites and it decreases for fainter satellite samples (see
\reffig{fig:theta_plane_vary_top_N}).
 \\[-.3cm]
 \item Dwarfs accreted in groups and along the richest filament have
infall directions that are more anisotropic than the full satellite
sample (see Figs \ref{fig:sky} and \ref{fig:theta_mul}). It suggests
that the filament that dominates the anisotropic accretion of matter,
and thus determines the halo orientation, is also the one that brings
both the most satellites falling in groups and the most satellites
overall.
\end{enumerate}

One of the goals of this paper was to understand what determines the
spatial and kinematic structures seen in the Galactic distribution of
satellites, the so-called MW disc of satellites. Motivated by previous
literature
\citep[e.g.][]{Libeskind2005,Libeskind2011,Li2008,Lovell2011}, we
checked if indeed enhanced group or filamentary accretion leads to a
larger amount of structure in the distribution of dwarf satellites
(see \reffig{fig:am-mul}). The accretion of very rich groups, which is
rare, does lead to flatter spatial distributions, but it does not
enhance the number of satellites with similar orbital poles. Such rich
groups typically arrive in massive haloes and thus their members can
have a large velocity dispersion, which can lead to different orbital
planes. MW-mass systems that accreted most of their satellites along a
single filament have the same average flattening and degree of planar
orbits of their $z=0$ satellites as the overall sample. If indeed accretion
along filaments is responsible for rotating planes of satellites, then our results suggest
that the connection between the MW disc of satellites and filamentary
accretion is not as simple as having the majority of satellites
accreted along one filament, and that the important factors might be
the spatial configuration and the characteristics of the filaments
surrounding the host halo.

\section*{Acknowledgements}
We thank the anonymous referee for detailed comments that have helped us improve the paper.
We also thank Alis Deason, Jie Wang, and Andrew Cooper for helpful discussions.
SS, MC and CSF were supported by the Science and Technology
Facilities Council [grant number ST/F001166/1, ST/I00162X/1, ST/P000451/1].
MC and CSF were also supported in part by ERC Advanced Investigator grant
COSMIWAY [grant number GA 267291]. RG acknowledges support by the DFG Research Centre SFB-881
`The Milky Way System' through project A1.
This work used the DiRAC Data Centric system at Durham University, operated by ICC on behalf of the STFC DiRAC HPC Facility (www.dirac.ac.uk). This equipment was funded by BIS National E-infrastructure capital grant ST/K00042X/1, STFC capital grant ST/H008519/1, and STFC DiRAC Operations grant ST/K003267/1 and Durham University. DiRAC is part of the National E-Infrastructure. We acknowledge PRACE for awarding us
access to the Curie machine based in France at TGCC, CEA,
Bruy\`eres-le-Ch\^atel. Some of the results in this paper have
used the \healpix{} package described in \citet{Gorski2005}.

\bibliographystyle{mnras}
\bibliography{bibliography}

\appendix
\section{Filament distribution}
\label{appendix:filament_distribution}
\begin{figure}
  \centering
  \plotone{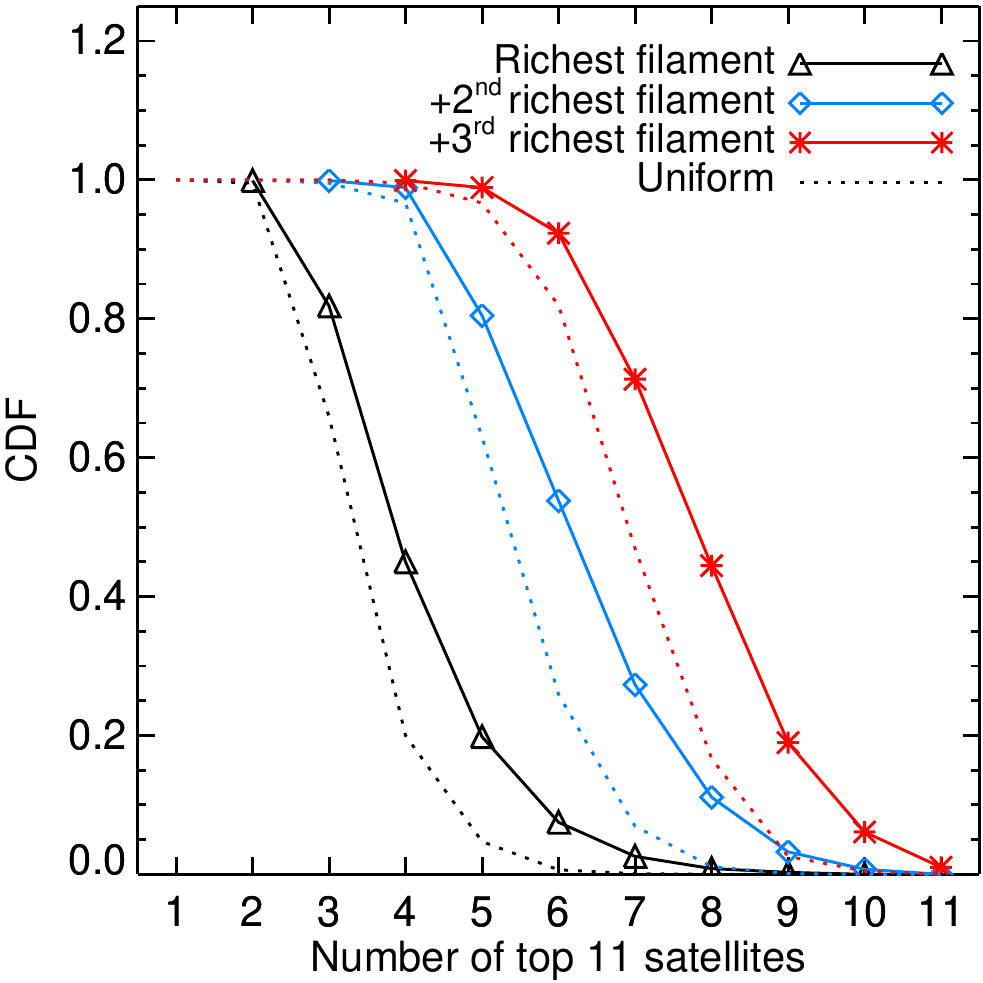}
  \caption{ The CDF of the number of satellites in the richest, the two richest and three richest filaments. The solid lines with symbols show the \eagle{} results while the dotted lines show the corresponding expectation if the satellite accretion directions are distributed isotropically on the sky.}
  \label{fig:filaments_isotropic}
\end{figure}
\begin{figure}
  \centering
  \plotone{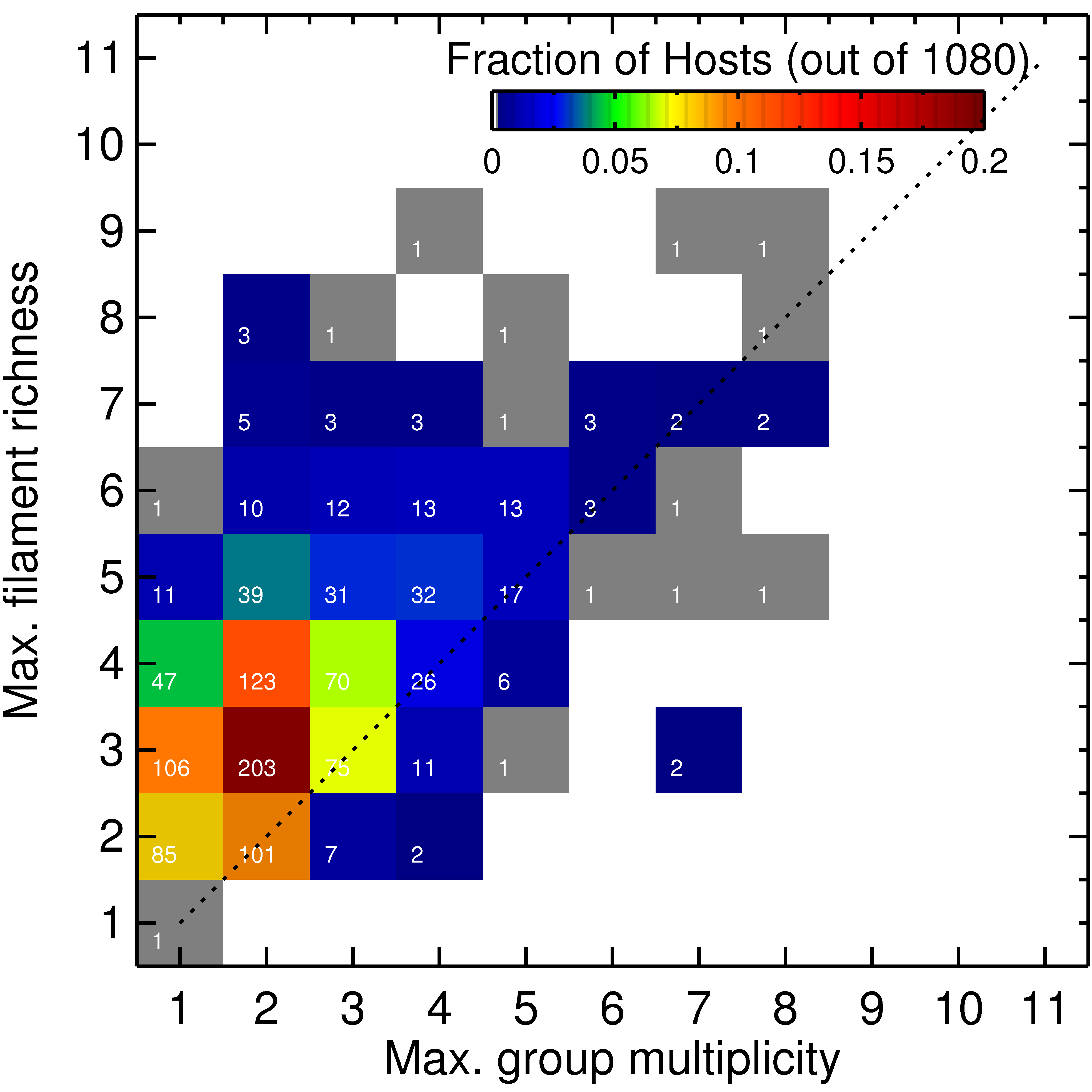}
  \caption{ The maximum filament richness as a function of the maximum group multiplicity for \eagle{} MW-mass haloes. Each square is coloured according to the fraction of the population that it contains and the numbers inside give the number of systems that contribute to that point. The diagonal line corresponds to filament richness and group multiplicity having the same value. For most cases, the richest filament has a higher richness than the richest group.}
  \label{fig:richness_correlation}
\end{figure}

To assess the degree of filamentary accretion of satellites, it is 
instructive to compare to the case when the satellite entry directions 
are distributed isotropically on the sky. This comparison is shown in
\reffig{fig:filaments_isotropic}, where we find that $\Lambda$CDM filaments 
have higher richness than in the isotropic case. For example, the top filament 
has a richness of 4 or more in 45\% of the hosts, while the isotropic accretion 
case results in a similarly rich top filament in only 20\% of systems.
Furthermore, the top three filaments bring seven or more satellites for 70\% 
of the MW-mass haloes, while the isotropic accretion case results in 
similarly rich top three filaments in only 45\% of systems.

In \reffig{fig:richness_correlation} we study the relation between 
the maximum filamentary richness and the maximum multiplicity of accretion for each 
of the \eagle{} MW-mass hosts. Whereas for the majority of host haloes the 
top filament has higher richness than the top group, in a significant 
fraction of systems (21\%) the top filament and the top group have the 
same richness. Thus, some of the richest filaments consist of satellites 
accreted in a single group, with no additional satellites falling in 
along the same direction. Furthermore, for a very small fraction of the \eagle{} 
systems (3\%) the top filament is less rich than the top group. This is 
due to the groups that have sizes on the sphere of the sky larger than 
a $30^\circ$ opening angle, which is the value used to define 
filaments. The large size
of some groups could be due to them living in massive hosts, which could
be the case for the very rich groups (e.g. group multiplicity $\gtrsim 5$), 
whereas the low-multiplicity groups are likely due to interloper members 
that are misidentified as being part of an extended FOF group.

\section{The connection between satellite planes and anisotropic accretion}
\label{sect:appendix_two}
\begin{figure}
 \plotone{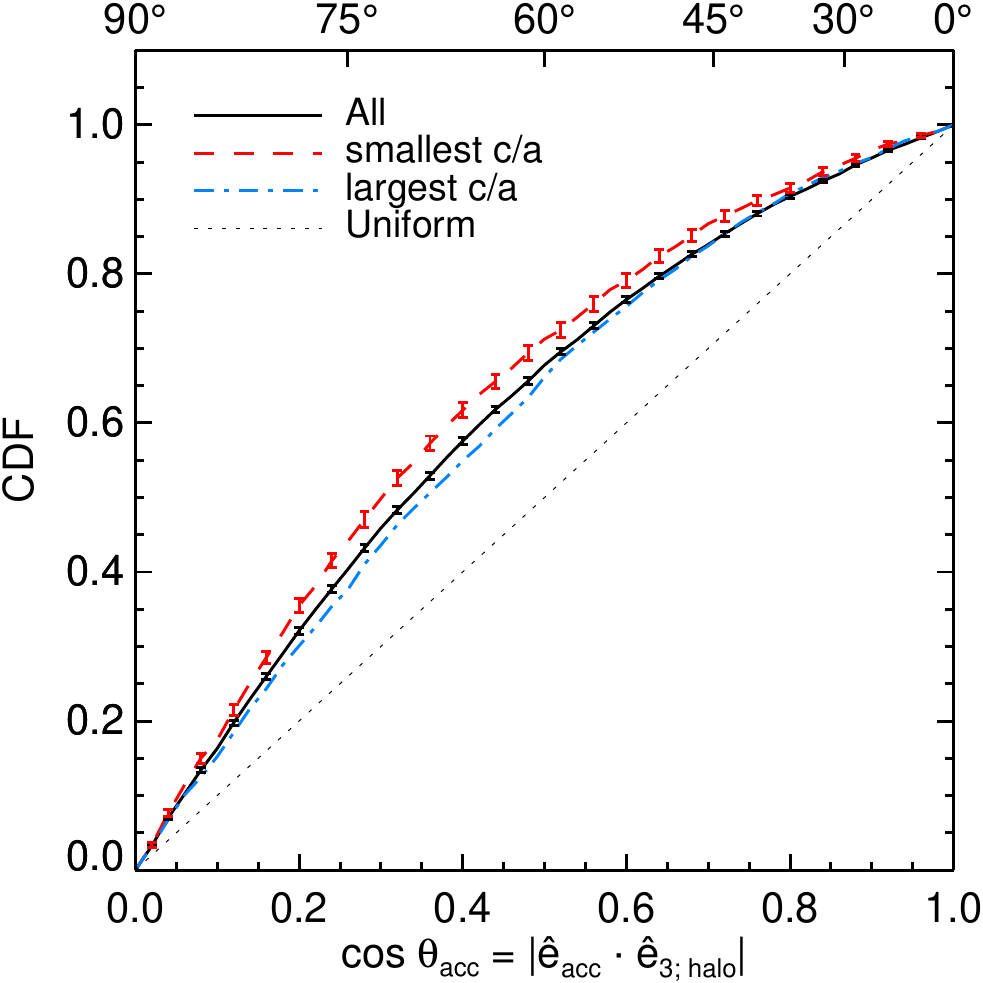}
 \caption{The CDF of the accretion misalignment angle, $\theta_{\rm
acc}$, between the entry points of the top 11 satellites and the shape
minor axis of the $z=0$ host halo. It shows the full sample (solid
black) and two subsamples consisting of the 20\% of \eagle{} host
haloes with the smallest (red dashed) and largest (blue dot-dashed)
$c/a$ values for the top 11 satellite distribution.
 }
 \label{fig:theta_ca}
\end{figure}

Here we study in more detail the factors that could explain the
flattening of the classical Galactic satellites. We split the \eagle{}
halo into two subsamples according to the $z=0$ flattening of the top 11
satellite system, as measured by the $c/a$ ratio (see Fig. 2 in
\citealt{Shao2016} for the $c/a$ distribution). We select the 20\%
of MW-mass haloes that have the thinnest ($c/a<0.33$) and the thickest
($c/a>0.57$) satellite distributions. \reffig{fig:theta_ca} shows the
accretion misalignment angle between the satellite entry points and
the shape minor axis of the $z=0$ DM halo for the two
subsamples. Present-day satellite systems that are thin are more
likely to have more anisotropic accretion, while the converse is true
for thicker satellite systems. Similarly, while not shown, we have
also studied the alignment between satellite entry points and the
minor axis of the present-day satellite system to find a similar
correlation: thinner $z=0$ satellite distributions correspond to more
anisotropic accretion.
\label{lastpage}

\end{document}